\documentclass[10pt,twocolumn,amsmath,amssymb,aps,pra,showpacs,longbibliography,superscriptaddress]{revtex4-1}
\usepackage[latin9]{inputenc}
\usepackage{amsmath}
\usepackage{amssymb}
\usepackage{graphicx}
\usepackage{esint}
\PassOptionsToPackage{normalem}{ulem}
\usepackage{ulem}

\makeatletter

\usepackage{amsfonts}
\usepackage{graphics}

\makeatother

\begin{document}
\title{Exact theory of the finite-temperature spectral function of Fermi
polarons with multiple particle-hole excitations: Diagrammatic theory
versus Chevy ansatz}
\author{Hui Hu}
\affiliation{Centre for Quantum Technology Theory, Swinburne University of Technology,
Melbourne 3122, Australia}
\author{Jia Wang}
\affiliation{Centre for Quantum Technology Theory, Swinburne University of Technology,
Melbourne 3122, Australia}
\author{Xia-Ji Liu}
\affiliation{Centre for Quantum Technology Theory, Swinburne University of Technology,
Melbourne 3122, Australia}
\date{\today}
\begin{abstract}
By using both diagrammatic theory and Chevy ansatz approach, we derive
an exact set of equations, which determines the spectral function
of Fermi polarons with multiple particle-hole excitations at nonzero
temperature. In the diagrammatic theory, we find out the complete
series of Feynman diagrams for the multi-particle vertex functions,
when the unregularized contact interaction strength becomes infinitesimal,
a typical situation occurring in two- or three- dimensional free space.
The latter Chevy ansatz approach is more widely applicable, allowing
a nonzero interaction strength. We clarify the equivalence of the
two approaches for an infinitesimal interaction strength and show
that the variational coefficients in the Chevy ansatz are precisely
the on-shell multi-particle vertex functions divided by an excitation
energy. Truncated to a particular order of particle-hole excitations,
our exact set of equations can be used to numerically calculate the
finite-temperature polaron spectral function, once the numerical singularities
in the equations are appropriately treated. As a concrete example,
we calculate the finite-temperature spectral function of Fermi polarons
in one-dimensional lattices, taking into account all the two-particle-hole
excitations. We show that the inclusion of two-particle-hole excitations
quantitatively improve the predictions on the polaron spectral function.
Our results provide a useful way to solve the challenge problem of
accurately predicting the finite-temperature spectral function of
Fermi polarons in three-dimensional free space. In addition, our clarification
of the complete set of Feynman diagrams for the multi-particle polaron
vertex functions may inspire the development of more accurate diagrammatic
theories of population-imbalanced strongly interacting Fermi gases,
beyond the conventional many-body $T$-matrix approximation.
\end{abstract}
\maketitle

\section{Introduction}

A single impurity interacts with a Fermi sea of ideal, non-interacting
fermions, which is the so-called Fermi polaron, is probably the simplest
quantum many-body system that one may imagine \citep{Alexandrov2010}.
In the limit of an infinitely heavy impurity, which, to a good approximation,
arises in the study of the Fermi edge singularity in x-ray spectra
\citep{Mahan1967}, a highly non-trivial exact solution exists \citep{TextbookMahan2000,Roulet1969,Nozieres1969a,Nozieres1969b}.
As described in detail by Mahan in his classical textbook \citep{TextbookMahan2000},
the development of such a seemingly simple exact solution is actually
not straightforward. In his first investigation of the Fermi edge
singularity, Mahan calculated various Feynman diagrams at the high
orders of the interaction strength and elucidated that it is inevitable
to include the vertex corrections \citep{Mahan1967}, beyond the many-body
$T$-matrix approximation that sums all the ladder diagrams. This
insightful observation motivated Nozières and colleagues to consider
the parquet diagrams \citep{Roulet1969} and eventually fully solve
the Fermi edge singularity problem with the parquet equation \citep{Nozieres1969a}.
Interestingly, only after this staggering journey of diagrammatic
calculations, which clarify the essence of the Fermi edge singularity,
the exact solution starts to emerge \citep{Nozieres1969b}. Most recently,
equipped with a new tool of the functional determinant approach \citep{Levitov1996},
this exact solution has been used to systematically analyze the quasiparticle
properties of heavy Fermi polarons \citep{Knap2012,Schmidt2018,Wang2022PRL,Wang2022PRA,Wang2023,Wang2023AB}.

Away from the heavy impurity limit, Fermi polarons turn out to be
notoriously difficult to handle, especially when the interparticle
interaction between the impurity and the Fermi sea becomes strong.
Fortunately, the recent rapid advances in the ultracold atomic research
open an entirely new avenue to experimentally explore the Fermi polaron
physics \citep{Knap2012,Schmidt2018,Massignan2014,Scazza2022}. By
using various spectroscopic tools, such as the radio-frequency spectroscopy
\citep{Schirotzek2009,Zhang2012,Zan2019}, Ramsey interferometry \citep{Cetina2016},
Rabi oscillation \citep{Scazza2017,Vivanco2024}, and Raman spectroscopy
\citep{Ness2020}, quasiparticle properties of Fermi polarons have
been quantitatively characterized. These results on the polaron energy,
effective mass and quasiparticle residue urgently require the development
of an accurate theoretical description of Fermi polarons, particularly
on the finite-temperature polaron spectral function, which underlies
the various spectroscopic measurements.

A fundamental theoretical framework of Fermi polarons is the so-called
Chevy ansatz, which was frequently used by condensed matter community
to investigate the Nagaoka ferromagnetism \citep{Nagaoka1966} a few
decades ago \citep{Shastry1990,Basile1990,vonderLinden1991,Cui2010},
but re-attracted great attention due to the seminal work by Chevy
in 2006 \citep{Chevy2006}. The Chevy ansatz is generally recognized
as a variational approach \citep{Chevy2006,Combescot2008,Giraud2010,Parish2013,Liu2019,Liu2022},
which is applicable at zero temperature. It successively takes into
account the shake-up process (i.e., the multi-particle-hole excitations)
of the Fermi sea due to the interaction with the impurity. Remarkably,
already at the first order with only one-particle-hole excitations,
the Chevy ansatz works exceptionally well for a strongly interacting
Fermi polaron in three-dimensional free space, where the $s$-wave
scattering length between the impurity and fermions diverges. It predicts
an accurate polaron energy for the ground-state attractive polaron
at zero temperature, with a relative error less than $2\%$, as benchmarked
by the ab-initio quantum Monte Carlo simulation \citep{Prokofev2008,Pessoa2021}.
Another useful theoretical framework of Fermi polarons is the diagrammatic
theory \citep{Combescot2007,Hu2018,Tajima2018,Mulkerin2019,Tajima2019,Hu2022a,Hu2022b,Hu2022c,Hu2023AB,Hu2023}.
Interestingly enough, for Fermi polarons in three-dimensional free
space, the Chevy ansatz with one-particle-hole excitations is shown
to be fully equivalent to the many-body $T$-matrix theory at zero
temperature \citep{Combescot2007}.

The two theoretical mainstreams, the Chevy ansatz with one-particle-hole
excitations and the many-body $T$-matrix theory, have now been widely
used to understand the spectroscopic observations of Fermi polarons
\citep{Massignan2014}. They show a great success in predicting the
ground-state attractive polaron energy \citep{Massignan2014}. However,
they fail to quantitatively explain the spectroscopic data of the
excited states of Fermi polarons at high energy, particularly at nonzero
temperature \citep{Tajima2019,Hu2022b,Hu2022c}. Other non-perturbative
approaches, such as the functional renormalization group \citep{Schmidt2011,vonMilczewski2024}
and quantum Monte Carlo \citep{Goulko2016}, have also been considered.
However, their comparisons with the experimental observations are
mostly restricted to the zero-temperature situation.

Theoretically, the failure of the first-order Chevy ansatz and many-body
$T$-matrix theory is easy to understand in the heavy polaron limit.
In the diagrammatic language, both approaches do not include the important
vertex corrections, which should become crucial, as the mass ratio
between the impurity and fermions increases \citep{Mahan1967,TextbookMahan2000}.
Therefore, to quantitatively understand the current spectroscopic
data, we must include the multi-particle-hole excitations of the Fermi
sea, both in the Chevy ansatz and in the diagrammatic theory. However,
to date, only the effect of two-particle-hole excitations on the attractive
polaron energy at zero temperature has been considered \citep{Combescot2008,Giraud2010,Parish2013}.

In this work, we formally derive an exact set of infinitely many equations
by using both the diagrammatic theory and Chevy ansatz, which can
be used to determine the finite-temperature spectral function of Fermi
polarons. Our set of equations accounts for arbitrary numbers of particle-hole
excitations. Truncated to a particular order $n$, the set of equations
encloses and thus provides an approximate theory with the inclusion
of $n$-particle-hole excitations of the Fermi sea. At the first order
(i.e., $n=1$), our theory reduces to the conventional many-body $T$-matrix
approximation \citep{Combescot2007,Hu2022b}. However, at higher orders
($n\geq2$), the theory gradually adds the non-trivial vertex corrections
and should become more accurate. 

A brief summary of our work is given in a companion short Letter \citep{ShortPRL2024}.
Here, we would like to highlight a few non-trivial points. First,
it is somehow unexpected that we are able to find out the complete
Feynman diagrams for the multi-particle vertex functions, which is
very rare for quantum many-particle systems. Certainly, this is a
peculiar property of Fermi polarons, since the single impurity is
only allowed to propagate forward in time \citep{Roulet1969}. Second,
although the first-order Chevy ansatz is shown to be equivalent to
the many-body $T$-matrix theory \citep{Combescot2007}, a general
relation between the key variables in the two approaches is not known.
Our work clearly reveals that, for an infinitesimal unregularized
interaction strength, the variational coefficients in the Chevy ansatz
precisely represent the on-shell multi-particle vertex functions in
the diagrammatic theory, up to a factor of the excitation energy.
This interesting relation demonstrates the powerfulness of the variational
approach, as the multi-particle vertex functions are generally impossible
to obtain in the diagrammatic theory, even for on-shell values. Finally,
despite the exactness of our equations, they suffer from numerical
singularities, when some excitation energies become resonant. As a
result, the variational coefficients or the on-shell multi-particle
vertex functions often vary dramatically. It is interesting to note
that, in the heavy impurity limit, the parquet equation used by Nozières
and colleagues back to 1960s also suffers from logarithmic singularities
\citep{Roulet1969}, which are tamed through analytic analysis \citep{Nozieres1969a}.
We are optimistic that some clever ideas might be developed in the
future analysis of our exact set of equations, with which the singularities
could be removed.

The rest of the paper is organized as follows. In the next section
(Sec. II), we outline the model Hamiltonian of Fermi polarons. In
Sec. III, we construct the diagrammatic theory for an infinitesimal
unregularized interaction strength. We explain in detail the structures
of diagrammatic representation of multi-particle vertex functions
and set up the exact set of equations for the polaron self-energy
and hence the polaron spectral function at nonzero temperature. In
Sec. IV, we turn to the Chevy ansatz approach with a general nonzero
interaction strength. We derivate the equations satisfied by the variational
coefficients, with the inclusion of arbitrary numbers of particle-hole
excitations. We also consider the specific case of a vanishingly small
interaction strength and simplify the equations for the variational
coefficients. In Sec. V, we show that the same exact set of equations
for the multi-particle vertex functions can be recovered, if we identify
the variational coefficients as the on-shell multi-particle vertex
functions divided by the excitation energy. In Sec. VI, we discuss
the singularities in the derived exact set of equations and propose
that, for Fermi polarons in lattices, the singularities might be removed
by introducing a finite broadening factor and then extrapolating it
to zero. We consider Fermi polarons in one-dimensional lattices as
a concrete example, and numerically determine the finite-temperature
polaron spectral function, with the inclusion of two-particle-hole
excitations. Finally, Sec. VII is devoted to the conclusions and outlooks.

\section{Model Hamiltonian}

We consider a population imbalanced spin-1/2 Fermi gas in free space
or within lattices. In the extremely imbalanced limit, i.e., one spin-down
fermion immersed in a Fermi sea of spin-up fermions, we treat the
spin-down fermion as the impurity \citep{Massignan2014,Schirotzek2009}.
The system is then well described by the model Hamiltonian, $\mathcal{H}=\mathcal{H}_{0}+\mathcal{H}_{\textrm{int}}$,
where
\begin{eqnarray}
\mathcal{H}_{0} & = & \sum_{\mathbf{k}}\varepsilon_{\mathbf{k}}c_{\mathbf{k}}^{\dagger}c_{\mathbf{k}}+\sum_{\mathbf{p}}\varepsilon_{\mathbf{p}}^{I}d_{\mathbf{p}}^{\dagger}d_{\mathbf{p}},\\
\mathcal{H}_{\textrm{int}} & = & U\sum_{\mathbf{KK}'\mathbf{QQ}'}\delta_{\mathbf{K}+\mathbf{Q},\mathbf{K}'+\mathbf{Q}'}c_{\mathbf{K}}^{\dagger}c_{\mathbf{K}'}d_{\mathbf{Q}}^{\dagger}d_{\mathbf{Q}'}.
\end{eqnarray}
For clarity, we have suppressed the system volume. Here, $c_{\mathbf{k}}^{\dagger}$
and $d_{\mathbf{p}}^{\dagger}$ are the creation field operators for
spin-up fermions and the impurity, with single-particle dispersion
relations $\varepsilon_{\mathbf{k}}$ and $\varepsilon_{\mathbf{p}}^{I}$,
respectively. In free space, $\varepsilon_{\mathbf{k}}=\hbar^{2}k^{2}/(2m)$
and $\varepsilon_{\mathbf{p}}^{I}=\hbar^{2}p^{2}/(2m_{I})$ and we
allow different masses $m$ and $m_{I}$ for spin-up fermions and
the impurity, respectively. In lattices, for example, in one-dimensional
lattices, we would instead have $\varepsilon_{\mathbf{k}}=-2t\cos k+2t$
and $\varepsilon_{\mathbf{p}}^{I}=-2t_{d}\cos p+2t_{d}$, where the
mass difference is represented by the different hopping strength of
spin-up fermions ($t$) and of the impurity ($t_{d}$) on the lattice.
In the interaction Hamiltonian $\mathcal{H}_{\textrm{int}}$, $U$
is the interaction strength of the zero-range contact interaction
potential (in free space) or of the on-site interaction (in lattices).
The Dirac delta function $\delta_{\mathbf{K}+\mathbf{Q},\mathbf{K}'+\mathbf{Q}'}$
ensures the momentum conservation during the interparticle collisions.

In two- or three-dimensional free space, it is well-known that the
$s$-wave contact interaction is not physical at high energy scale
(i.e., above a momentum cut-off $\Lambda$) and we need to regularize
the associated ultraviolet divergence \citep{Hu2006,He2015}. For
example, in three dimensions we should replace the running interaction
strength $U(\Lambda)$ with a given $s$-wave scattering length $a$, 

\begin{equation}
\frac{1}{U\left(\Lambda\right)}=\frac{m_{r}}{2\pi\hbar^{2}a}-\sum_{\mathbf{\left|p\right|}<\Lambda}\frac{2m_{r}}{\hbar^{2}\mathbf{p}^{2}},\label{eq: Ureg}
\end{equation}
where $m_{r}\equiv mm_{I}/(m+m_{I})$ is the reduced mass. It is easy
to see, the interaction strength $U(\Lambda)\rightarrow0^{-}$ becomes
infinitesimal, when we tune the cut-off $\Lambda$ to be infinitely
large. This is exactly the case that we will consider in the next
section of the diagrammatic theory. In contrast, in lattices, the
interaction strength takes a finite value, either attractive or positive. 

\section{The diagrammatic theory}

\begin{figure}
\begin{centering}
\includegraphics[width=0.35\textwidth]{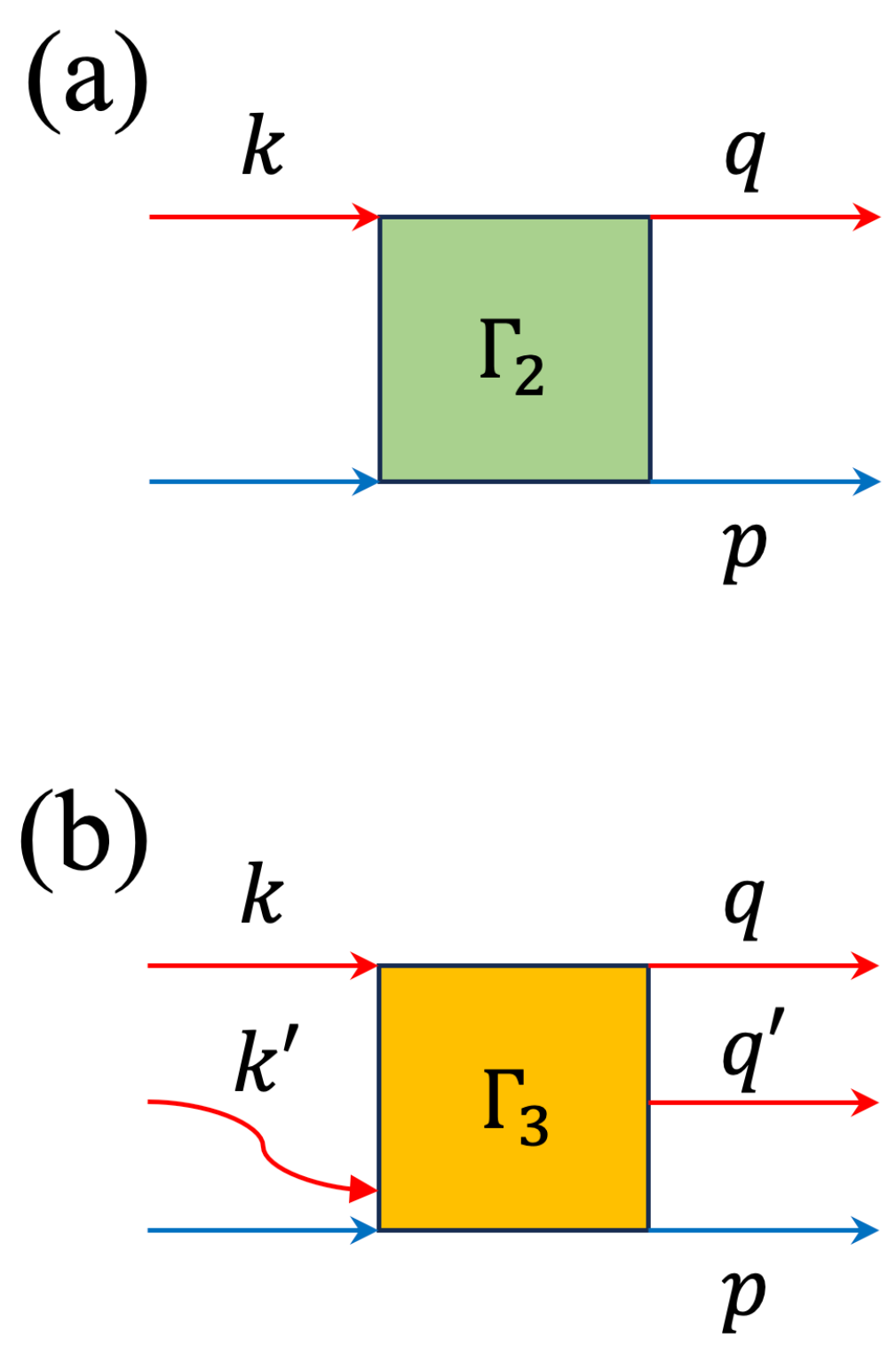}
\par\end{centering}
\caption{\label{fig1: VertexFunctions} The fundamental diagrams of vertex
functions: the (two-particle) vertex function $\Gamma_{2}(k;p,q)$
(a) and the three-particle vertex function $\Gamma_{3}(kk';p,qq')$
(b). Here, $p$, $k$, $k'$, $q$ and $q'$ are four-momenta. For
example, the four-momentum $p=(\mathbf{p},i\omega_{p})$ includes
both the spatial momentum $\mathbf{p}$ and the fermionic Matsubara
momentum $i\omega_{p}$, where $\omega_{p}=(2m_{p}+1)\pi k_{B}T$
with integer $m_{p}$ at the temperature $T$. For $\Gamma_{3}$ with
two spin-up fermions (see the two red lines) and one-spin down fermion
(i.e., the impurity, as indicated by a blue line), we let the spin-up
fermion with moment $k'$ interacts first with the impurity. On the
other hand, we do not care about the order of the two out-going spin-up
fermion lines. However, the Fermi statistics requires the antisymmetrization
of $\Gamma_{3}$ upon exchanging $q$ and $q'$, i.e., $\Gamma_{3}(kk';p,q'q)=-\Gamma_{3}(kk';p,qq')$.}
\end{figure}

In the diagrammatic theory, the fundamental quantities that we need
to look after are the vertex function $\Gamma_{2}(k;p,q)$ (see Fig.
\ref{fig1: VertexFunctions}(a)) and the multi-particle (i.e., $n+1$
particles with $n\geq2$) vertex functions \citep{TextbookMahan2000,Giraud2010,TextbookAGD1975},
\begin{equation}
\Gamma_{n+1}\left(\{k_{l}\};p,\{q_{l}\}\right)\equiv\Gamma_{n+1}\left(k_{1}\cdots k_{n};p,q_{1}\cdots q_{n}\right),
\end{equation}
which describes the scatterings among $n$ spin-up fermions in the
Fermi sea and the impurity. These in-medium scatterings occur in the
presence of the Fermi sea, so more fermions other than the $n$ spin-up
fermions may participate at the intermediate stages. The momentum
$k_{l}\equiv(\mathbf{k}_{l},i\omega_{l})$, where $l=1,\cdots,n$,
is the four-momentum of the $l$-th incoming fermion line, involving
both the spatial momentum $\mathbf{k}_{l}$ and the fermionic Matsubara
frequency $\omega_{l}\equiv(2m_{l}+1)\pi k_{B}T$ with an integer
$m_{l}$ at the temperature $T$ \citep{TextbookMahan2000,TextbookAGD1975}.
In a similar way, we use the momentum $q_{l}$ to denote the four-momentum
of the $l$-th out-going fermion line. In addition, we use $p=(\mathbf{p},i\omega_{p})$
to label the four-momentum of the out-going impurity line. For small
values of $n$, i.e., $n=2$ or $n=3$, for convenience we also use
$k\equiv k_{1}$, $k'=k_{2}$, $k''=k_{3}$, and so on. The similar
notations hold for the out-going momentum $q_{l}$; see, for example,
Fig. \ref{fig1: VertexFunctions}(b) for the three-particle vertex
function $\Gamma_{3}(kk';p,qq')$.

\begin{figure*}[t]
\begin{centering}
\includegraphics[width=1\textwidth]{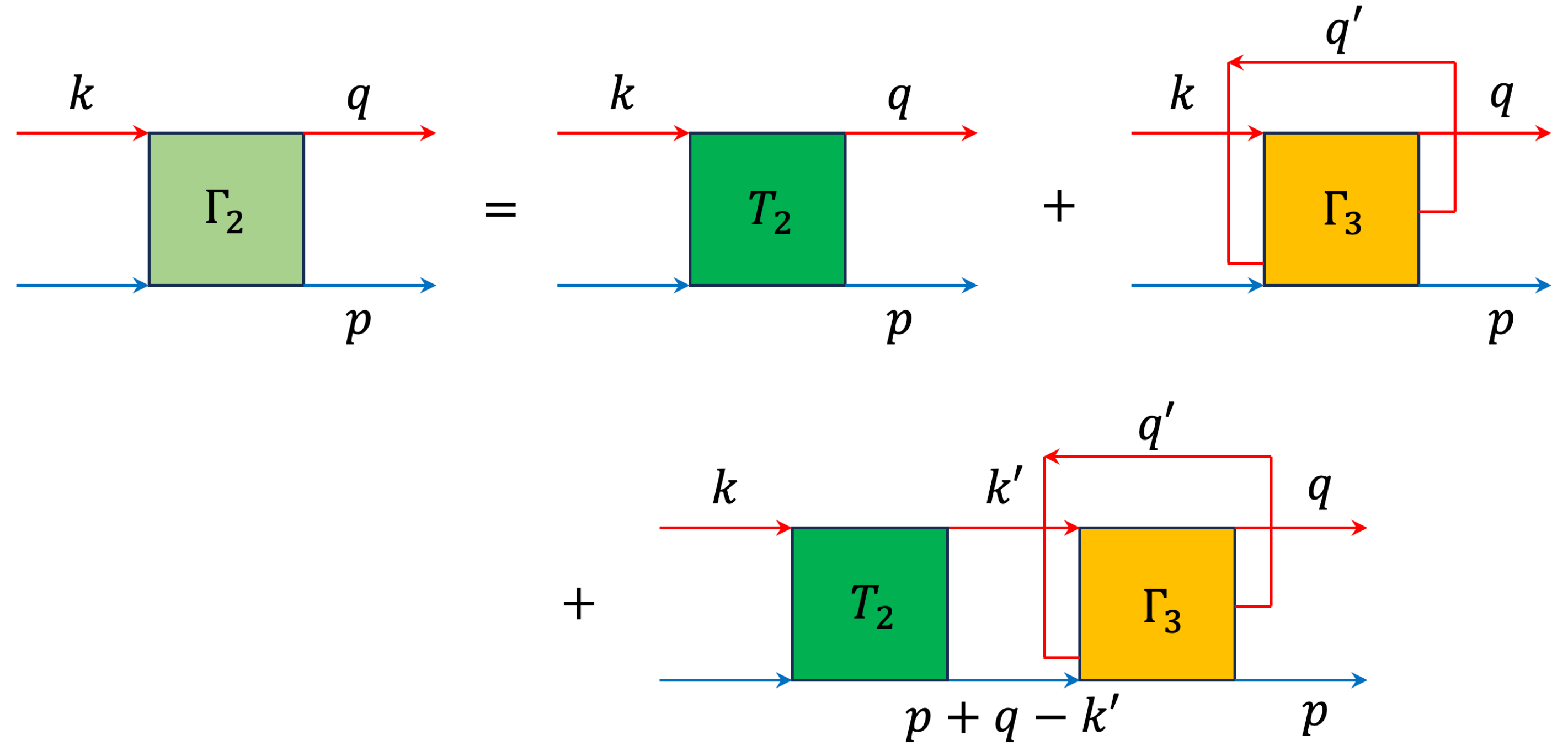}
\par\end{centering}
\caption{\label{fig2: Gamma2} The diagrammatic representation of the vertex
function $\Gamma_{2}(k;p,q)$. It includes the many-body $T$-matrix
$T_{2}$ that represents the summation of infinitely many ladder diagrams.
Any other higher order diagrams can be drawn in two ways, both of
which involve the three-particle vertex function $\Gamma_{3}$.}
\end{figure*}

For $\Gamma_{n+1}(k_{1}\cdots k_{n};p,q_{1}\cdots q_{n})$ with $n\geq2$,
the multi-particle vertex functions with more than one incoming fermion
line, we may have a freedom to pick up the line that interacts first
with the impurity. As a convention, we will always select the last
incoming fermion line with the momentum $k_{n}$ \citep{Giraud2010}.
This convention is useful for the contact interaction, where the many-body
$T$-matrix $T_{2}$ depends only on the total momentum. As we shall
see, the information of $k_{n}$ is then lost and the $(n+1)$-particle
vertex function becomes independent on $k_{n}$. We note also that,
following the Fermi statistics, the remaining $(n-1)$ incoming fermion
lines should be indistinguishable. As a result, the multi-particle
vertex function would acquire a minus sign, if we exchange any two
momenta among $k_{1}$, $\cdots$, $k_{n-1}$. The same antisymmetrization
requirement holds, upon exchanging any two out-going momenta among
$q_{1}$, $\cdots$, $q_{n}$. 

\subsection{Vertex function $\Gamma_{2}$}

Let us first consider the diagrammatic representation of the vertex
function $\Gamma_{2}(k;p,q)$ \citep{Giraud2010}. As shown in Fig.
\ref{fig2: Gamma2}, we may identify three contributions. The first
contribution is simply the familiar may-body $T$-matrix $T_{2}(p+q)$
that takes into account the successive scatterings between a spin-up
fermion and the impurity (i.e., ladder diagrams) \citep{Combescot2007,Hu2022b},
without the involvement of any other fermions in the Fermi sea. It
depends on the argument $p+q$ only, since we consider a zero-range
contact interaction for the impurity-fermion scattering. 

\begin{figure*}
\begin{centering}
\includegraphics[width=1\textwidth]{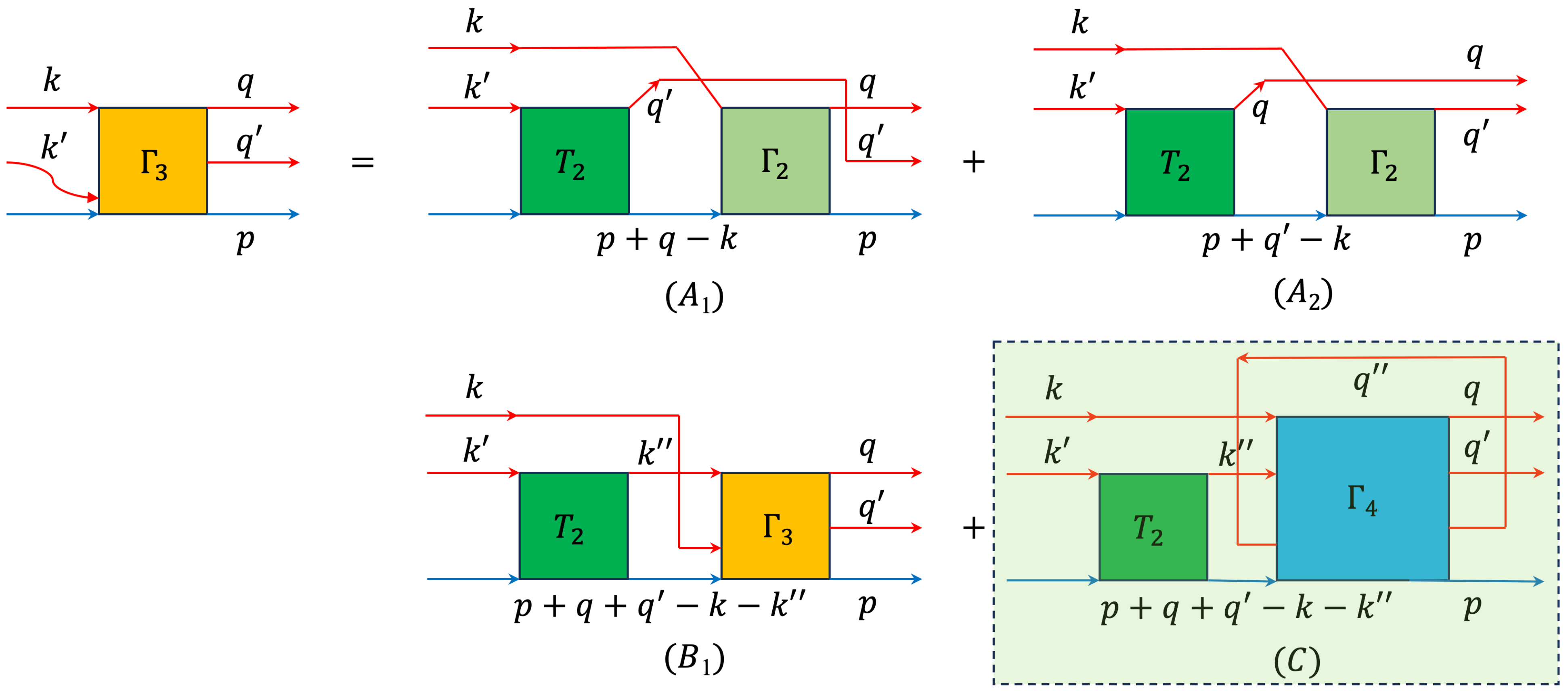}
\par\end{centering}
\caption{\label{fig3: Gamma3} The diagrammatic representation of the three-particle
vertex function $\Gamma_{3}(kk';p,qq')$. All the diagrams can be
categorized into three different types. The $A$-type diagrams, $A_{1}$
and $A_{2}$, involve a lower order vertex function $\Gamma_{2}$.
The $B$-type diagram, $B_{1}$, has the vertex function at the same
order. Finally, the $C$-type diagram consists of a higher order vertex
function $\Gamma_{4}$. Due to the requirement that the fermion line
with the momentum $k'$ interacts first with the impurity line, all
the three-type diagrams have a building block of the many-body $T$-matrix
$T_{2}$.}
\end{figure*}

The other two contributions necessarily involve the three-particle
vertex function, as we can see from the second and third diagrams
on the right-hand-side of Fig. \ref{fig2: Gamma2}. For the second
diagram, we wind one out-going fermion line (with the momentum $q'$)
back and connect it to the incoming fermion line with the momentum
$k'$, so $k'=q'$. It does not matter to wind the out-going $q$
or $q'$ line, since these two lines are chosen to be antisymmetric
and therefore we would obtain the exactly same contribution. But,
why we connect the out-going line with $q'$ to the incoming line
with the momentum $k'$? Logically, we could also connect it to the
upper incoming line with the momentum $k$. This way of connection
is actually realized in the next third diagram, as shown at the bottom
of Fig. \ref{fig2: Gamma2}. There, the first diagram ($T_{2}$) and
the second diagram that we have already been considered naturally
connect (see, for example, the diagram $B_{1}$ in Fig. \ref{fig3: Gamma3},
and imagine to connect the $q'$-line with the $k$-line).

\begin{figure}
\begin{centering}
\includegraphics[width=0.5\textwidth]{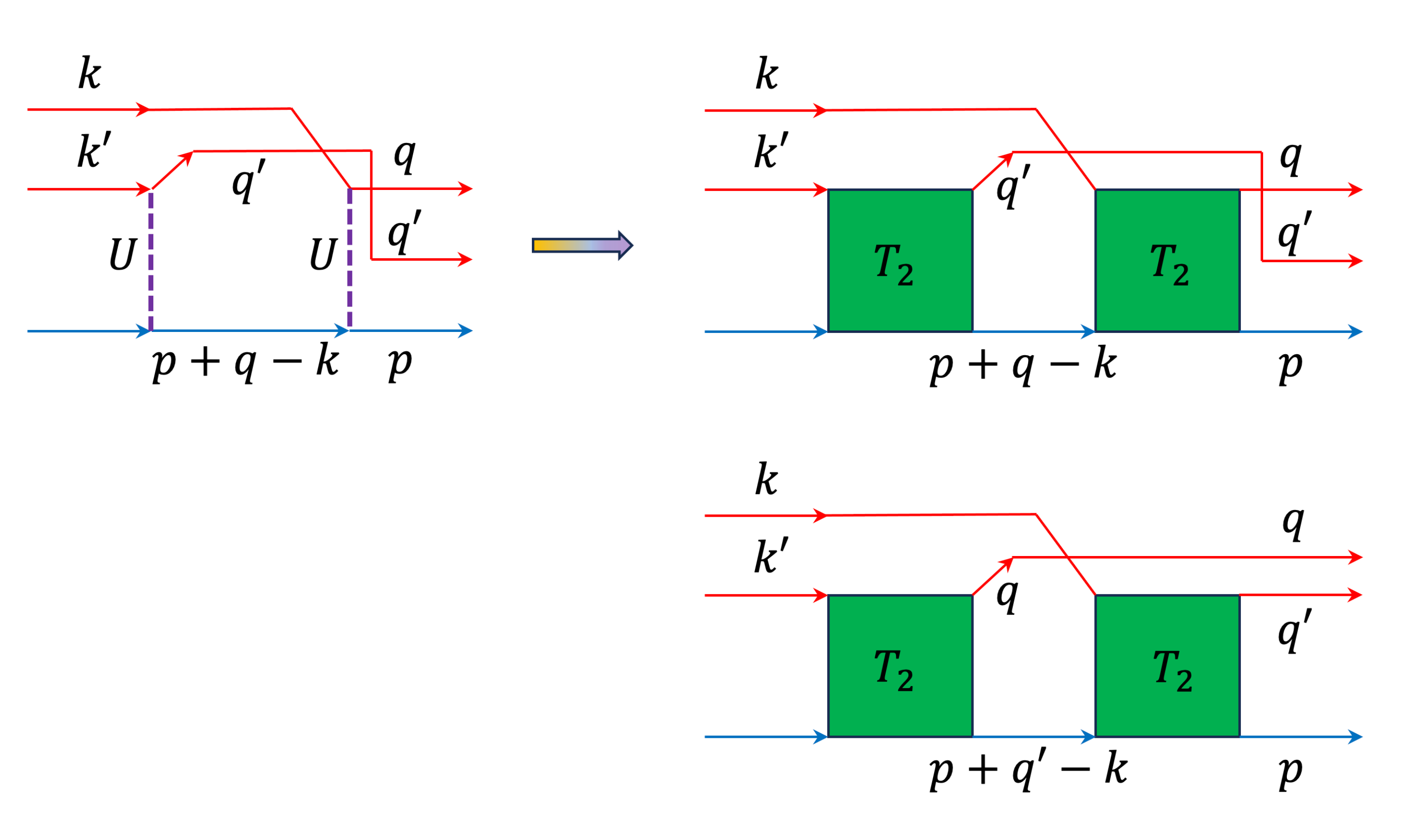}
\par\end{centering}
\caption{\label{fig4: T3BuildingBlocks} Left panel: A diagram of the three-particle
vertex function $\Gamma_{3}(kk';p,qq')$ at the lowest order of the
interaction strength, $U^{2}G_{0\downarrow}(p+q-k)$. Right panel:
The dashed interaction line in this diagram could be replaced with
the many-body $T$-matrix $T_{2}$, so we end up with a diagram at
the top right corner. In addition, we may exchange the two out-going
lines for spin-up fermions. This leads to the diagram at the bottom
right corner. The two diagrams at the right hand are covered by the
$A_{1}$ diagram and the $A_{2}$ diagram in Fig. \ref{fig3: Gamma3}.}
\end{figure}

\begin{figure}
\begin{centering}
\includegraphics[width=0.5\textwidth]{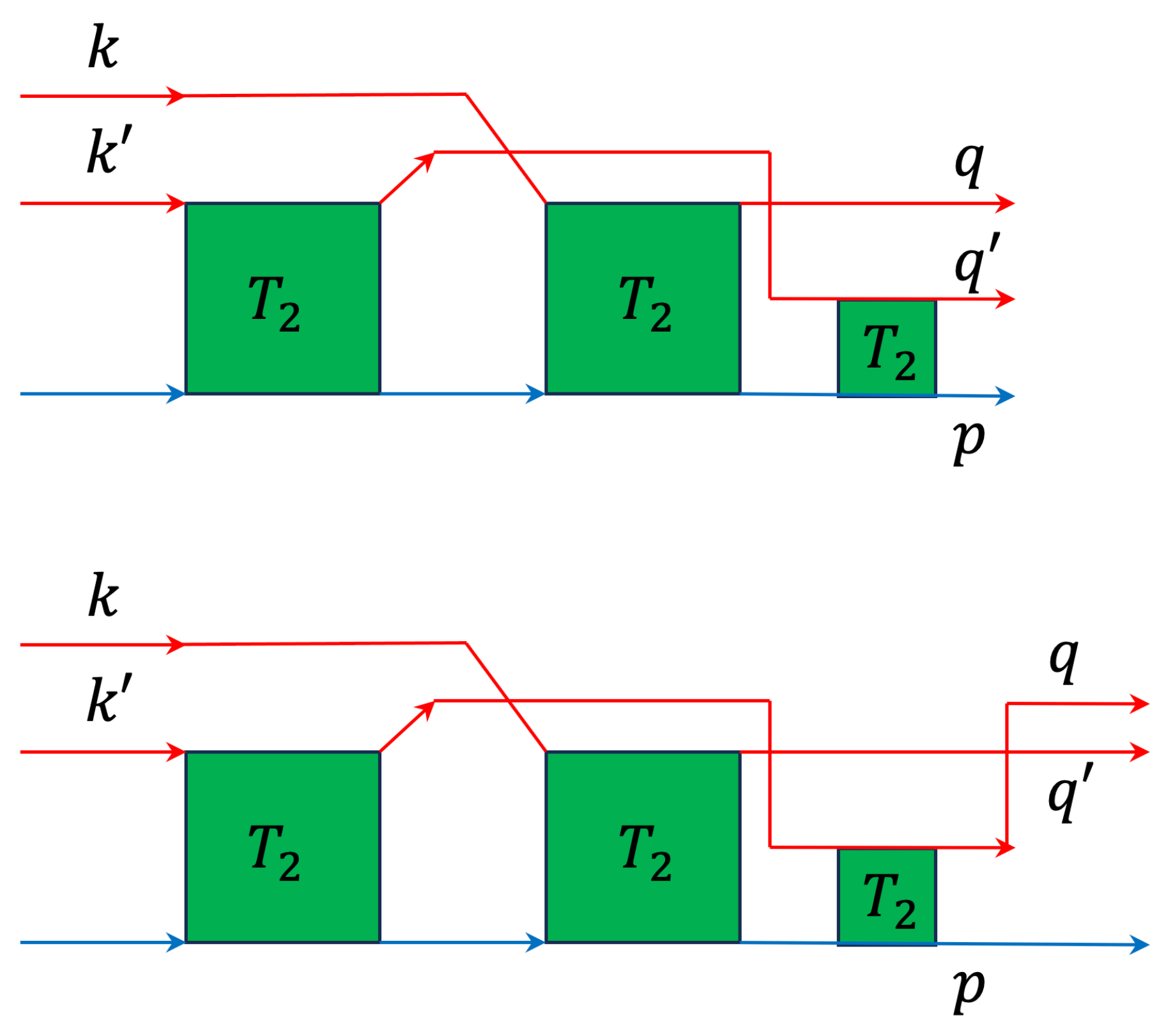}
\par\end{centering}
\caption{\label{fig5: T3MoreDiagrams} Two examples of higher-order diagrams
for the three-particle vertex function $\Gamma_{3}(kk';p,qq')$. As
in the right panel of Fig. \ref{fig4: T3BuildingBlocks}, the diagram
at the bottom is obtained by exchanging (i.e., anti-symmetrizing)
the two out-going fermion lines of the top diagram.}
\end{figure}

Now, we would like to claim that we exhaust all the possibilities
for constructing the vertex function $\Gamma_{2}(k;p,q)$. One may
wonder that for the last diagram in Fig. \ref{fig2: Gamma2}, we can
obtain a new contribution if we exchange the two blocks of $T_{2}$
and $\Gamma_{3}$. It is actually not true. This is because the three-particle
vertex function itself involves a lot of diagrams. If we attach the
many-body $T$-matrix $T_{2}$ to the right-hand-side of the three-particle
vertex function $\Gamma_{3}$, what happens is that the $T_{2}$ block
will be naturally absorbed into the $\Gamma_{3}$ block, giving a
contribution that we have already taken into account. Moreover, there
are no new contributions related to higher-order vertex functions
such as $\Gamma_{4}$, $\Gamma_{5}$ and so on, as their appearance
is implicitly included in the three-particle vertex function and we
need to avoid the double counting of diagrams. To see these, let us
carefully examine the diagrammatic representation of the three-particle
vertex function $\Gamma_{3}(kk';p,qq')$, which is given in Fig. \ref{fig3: Gamma3}.

\begin{figure*}
\begin{centering}
\includegraphics[width=1\textwidth]{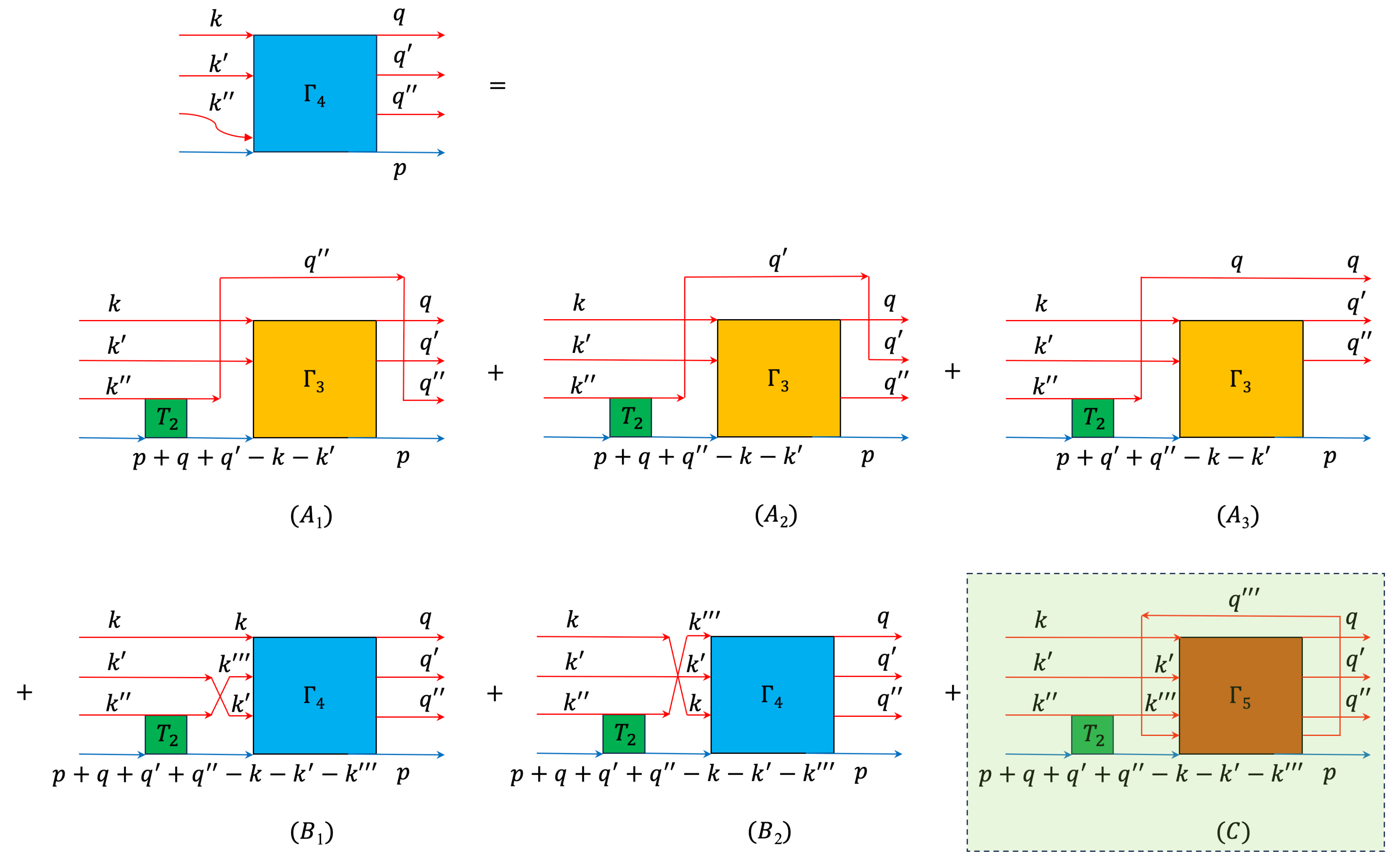}
\par\end{centering}
\caption{\label{fig6: Gamma4} The diagrammatic representation of the four-particle
vertex function $\Gamma_{4}(kk'k'';p,qq'q'')$. As in Fig. \ref{fig3: Gamma3},
all the diagrams can be categorized into three different types. The
$A$-type diagrams ($A_{1}$, $A_{2}$ and $A_{3}$), involve a lower
order vertex function $\Gamma_{3}$. The $B$-type diagram ($B_{1}$
and $B_{2}$) have the vertex function at the same order. Finally,
the $C$-type diagram consists of a higher order vertex function $\Gamma_{5}$.}
\end{figure*}

\subsection{Three-particle vertex function $\Gamma_{3}$}

There are four diagrams contributed to $\Gamma_{3}(kk';p,qq')$, which
might be categorized into three different types. To understand the
first two contributions in Fig. \ref{fig3: Gamma3}($A_{1}$) and
Fig. \ref{fig3: Gamma3}($A_{2}$), which are referred to as the $A$-type
diagrams, let us recall one of the lowest order diagrams of $\Gamma_{3}$,
measured in terms of the impurity-fermion interaction strength $U$,
as given in the left panel of Fig. \ref{fig4: T3BuildingBlocks}.
As we emphasized earlier, since we use the zero-range contact interaction,
the interaction strength $U$ effectively scales to zero when we increase
the cut-off momentum. To have a nonzero contribution, we then need
to replace the bare interaction line (i.e., the dashed line) in the
left panel of Fig. \ref{fig4: T3BuildingBlocks} everywhere by the
many-body $T$-matrix $T_{2}$. This replacement leads to the building
blocks of the three-particle vertex function $\Gamma_{3}$, as illustrated
in the right panel of Fig. \ref{fig4: T3BuildingBlocks}. Of course,
we may continue to consider higher order scattering processes, with
the two simplest examples illustrated in Fig. \ref{fig5: T3MoreDiagrams}.
It is not difficult to see that, some of the higher order scattering
processes could be easily taken into account, by simply replacing
in the two building blocks (i.e., the right panel of Fig. \ref{fig4: T3BuildingBlocks})
the second $T_{2}$ with the vertex function $\Gamma_{2}$. This replacement
leads to the two $A$-type diagrams, $A_{1}$ and $A_{2}$, in Fig.
\ref{fig3: Gamma3}.

This replacement, however, does not exhaust all the possibilities
for constructing $\Gamma_{3}$. Indeed, it is readily seen that in
the building blocks, we may replace the second $T_{2}$ by the three-particle
vertex function $\Gamma_{3}$ itself, and then connect the out-going
fermion line of the first $T_{2}$ to the first incoming fermion line
of $\Gamma_{3}$. This gives rise to the $B$-type diagram, $B_{1}$,
in Fig. \ref{fig3: Gamma3}. 

Finally, similar to what we have observed in constructing the vertex
function $\Gamma_{2}(k;p,q)$, we would have contributions to $\Gamma_{3}(kk';p,qq')$
that involves a higher order vertex function $\Gamma_{4}$. This is
given in the last $C$-type diagram in Fig. \ref{fig3: Gamma3}.

In all the diagrammatic contributions to $\Gamma_{3}(kk';p,qq')$,
we see clearly the involvement of the many-body $T$-matrix $T_{2}$.
It simply follows our definition of the multi-particle vertex functions,
as we impose the constraint that the last incoming fermion line must
interact first with the incoming impurity line. We will see that the
higher order vertex functions such as $\Gamma_{4}$ share the same
feature.

\subsection{Higher order vertex functions}

Actually, for all the higher order vertex functions, we have a stronger
conclusion that they all can be categorized into the similar three
different types. In Fig. \ref{fig6: Gamma4}, we show the diagrammatic
representation of the four-particle vertex function as an example
and explain the general rules to draw the three kinds of diagrams
for the $(n+1)$-particle vertex function $\Gamma_{n+1}$.

First, the $A$-type diagrams consist of a many-body $T$-matrix $T_{2}$
and a lower order $n$-particle vertex function $\Gamma_{n}$. The
out-going fermion line of $T_{2}$ is one of the out-going fermion
line of the whole diagram; see, for example, the $A_{1}$, $A_{2}$
and $A_{3}$ diagrams in Fig. \ref{fig6: Gamma4}. Thus, we have $n$
choices to place the out-going fermion line of $T_{2}$, which leads
to $n$ type-$A$ diagrams, listed as $A_{1}$, $\cdots$, $A_{n}$.

Second, the $B$-type diagrams involve a many-body $T$-matrix $T_{2}$
and the same order $(n+1)$-particle vertex function $\Gamma_{n+1}$.
The out-going fermion line of $T_{2}$ needs to connect with one of
the $(n-1)$ incoming fermion lines of $\Gamma_{n+1}$, apart from
the last incoming fermion line. Otherwise, we will obtain $\Gamma_{n+1}$
itself. Therefore, in total there are $(n-1)$ type-$B$ diagrams,
listed as $B_{1}$, $\cdots$, $B_{n-1}$.

The final $C$-type diagram is constructed by connecting a many-body
$T$-matrix $T_{2}$ to a higher order $(n+2)$-particle vertex function
$\Gamma_{n+2}$. It is not difficult to see that we only have one
possibility: we need to connect the last out-going fermion line of
$\Gamma_{n+2}$ to the last incoming fermion line of $\Gamma_{n+2}$,
owing to the antisymmetrization of $\Gamma_{n+2}$.

\subsection{Explicit expressions of various vertex functions}

The determination of the complete set of the diagrams for each multi-particle
vertex function allows us to directly write down its expression, in
terms of the Green functions of the spin-up fermions $G_{0\uparrow}$
and of the impurity Green function $G_{0\downarrow}$, and of the
multi-particle vertex functions themselves. However, specific attention
should be paid to the overall sign of each diagram, as we shall explain
in detail.

For example, let us write down the expression for the vertex function
$\Gamma_{2}(k;p,q)$, following its diagrammatic representation in
Fig. \ref{fig2: Gamma2},\begin{widetext}

\begin{equation}
\Gamma_{2}\left(k;p,q\right)=T_{2}\left(p+q\right)+\sum_{q'}G_{0\uparrow}\left(q'\right)\Gamma_{3}\left(k;p,qq'\right)-\sum_{k'q'}T_{2}\left(p+q\right)G_{0\uparrow}\left(k'\right)G_{0\downarrow}\left(p+q-k'\right)G_{0\uparrow}\left(q'\right)\Gamma_{3}\left(k';p,qq'\right).\label{eq: Gamma2}
\end{equation}
\end{widetext}Here, by default the summation $\sum_{k}$ (or $\sum_{q}$)
is understood as $k_{B}T\sum_{i\omega_{k}}\sum_{\mathbf{k}}$ (or
$k_{B}T\sum_{i\omega_{q}}\sum_{\mathbf{q}}$) for the four-momentum
$k\equiv(\mathbf{k},i\omega_{k})$ (or $q=(\mathbf{q},i\omega_{q})$)
\citep{TextbookMahan2000,TextbookAGD1975}. Each term on the right-hand-side
of Eq. (\ref{eq: Gamma2}) corresponds to a diagram in Fig. \ref{fig2: Gamma2}.
We note that, in the expression of $\Gamma_{3}\left(kk';p,qq'\right)$,
we have suppressed the argument of $k'$, as $\Gamma_{3}$ is independent
on it. 

The second term comes with a positive sign, although there is a Fermi
loop in the corresponding diagram (i.e., the second diagram on the
right-hand-side of Fig. \ref{fig2: Gamma2}), which contributes a
minus sign \citep{TextbookMahan2000,TextbookAGD1975}. This minus
sign is actually cancelled by the minus sign we wish to assign to
the three-particle vertex function $\Gamma_{3}$. Indeed, here we
encounter a potential ambiguity for the sign of $\Gamma_{3}$. Let
us recall the two building blocks of $\Gamma_{3}$, as illustrated
in the right part of Fig. \ref{fig4: T3BuildingBlocks}. These two
diagrams look very similar and differ only by exchanging the two out-going
fermion lines. Of course, this exchange means a sign difference. But,
which diagram has the positive sign? We naturally assume that the
upper building block of $\Gamma_{3}$ has the positive sign and write
down in an abbreviated form $\tilde{\Gamma}_{3}=T_{2}G_{0\downarrow}T_{2}$,
where the tilde on $\Gamma_{3}$ stands for the contribution from
the upper building block to $\Gamma_{3}$. Now, let us insert this
upper building block of $\Gamma_{3}$ into Fig. \ref{fig2: Gamma2}.
Following the standard rules, we may write down its contribution to
$\Gamma_{2}(k;p,q)$, i.e., $(-1)G_{0\uparrow}(-1)T_{2}G_{0\downarrow}T_{2}=(-1)G_{0\uparrow}(-1)\tilde{\Gamma}_{3}$,
again in an abbreviated form. Here, the first minus sign comes from
the Fermi loop and the second minus sign is because the diagram appears
to be an order higher in the interaction line (or in $T_{2}$), when
we replace $\Gamma_{3}$ with the upper building block of $\Gamma_{3}$.
In brief, the three-particle vertex function $\Gamma_{3}$ acquires
a minus sign, if we fix the sign convention for its two building blocks
given in the right panel of Fig. \ref{fig4: T3BuildingBlocks}. Following
such a convention, we can easily understand the minus sign appearing
in the third term of Eq. (\ref{eq: Gamma2}), as the corresponding
diagram is an order higher in $T_{2}$.

For the expression of the three-particle vertex function $\Gamma_{3}$,
according to Fig. \ref{fig3: Gamma3} we may formally write down,\begin{widetext}
\begin{eqnarray}
\frac{\Gamma_{3}\left(k;p,qq'\right)}{T_{2}\left(p+q+q'-k\right)} & = & +G_{0\downarrow}\left(p+q-k\right)\Gamma_{2}\left(k;p,q\right)-G_{0\downarrow}\left(p+q'-k\right)\Gamma_{2}\left(k;p,q'\right)\nonumber \\
 &  & +\sum_{k''}G_{0\uparrow}\left(k''\right)G_{0\downarrow}\left(p+q+q'-k-k''\right)\Gamma_{3}\left(k'';p,qq'\right)\nonumber \\
 &  & -\sum_{k''q''}G_{0\uparrow}\left(k''\right)G_{0\downarrow}\left(p+q+q'-k-k''\right)G_{0\uparrow}\left(q''\right)\Gamma_{4}\left(kk'';p,qq'q''\right),\label{eq: Gamma3}
\end{eqnarray}
\end{widetext}where the four terms on the right-hand-side of the
equations comes from the diagrams $A_{1}$, $A_{2}$, $B_{1}$ and
$C$, respectively. The signs of the first two terms have already
been discussed, since the diagrams $A_{1}$ and $A_{2}$ have the
exactly same topology as the two building blocks of $\Gamma_{3}$.
More generally, we will always fix the sign of the $A_{1}$-diagram
for any multi-particle vertex functions to be positive. The sign of
the $A_{j}$-diagram ($j=1,\cdots,n$) will then be $(-1)^{j-1}$.
Following this convention, it is not difficult to see that we need
to assign a minus sign to $\Gamma_{n+1}$ in relative to $\Gamma_{n}$.
As a result, all the $C$-type diagrams come with a negative sign.

It is a bit tricky to determine sign of the $B_{1}$ diagram in Fig.
\ref{fig3: Gamma3}. A convenient way is to compare the $B_{1}$ diagram
with the $A_{2}$ diagram. These two diagrams have one feature in
common: if we connect the out-going $q'$-line with the incoming $k'$
line, no Fermi loop is created. Thus, naïvely, we anticipate that
topologically the $B_{1}$ diagram will have the same sign as the
$A_{2}$ diagram. However, as $B_{1}$ involves $\Gamma_{3}$ instead
of $\Gamma_{2}$, an additional sign appears. This additional sign
will cancel the minus sign of the $A_{2}$ diagram. In the end, we
find a positive sign for the $B_{1}$ diagram. In the general case
for any multi-particle vertex functions $\Gamma_{n+1}$, we might
see that the sign of the $B_{i}$-type diagram ($i=1,\cdots,n-1$)
is always positive, in order to satisfy the requirement that there
should be a sign difference if we exchange any two momenta in $\{k_{l}\}_{l\neq n}$.

It is easy to check that the right-hand-side of Eq. (\ref{eq: Gamma3})
does not contain the momentum $k'$, justifying our previous statement
that $\Gamma_{n+1}(k_{1}\cdots k_{n};p,q_{1}\cdots q_{n})$ does not
depend on the last incoming momentum $k_{n}$. 

We are now ready to write down the general expression for the multi-particle
vertex function $\Gamma_{n+1}$ ($n\geq2$),
\begin{equation}
\frac{\Gamma_{n+1}\left(\{k_{l}\}_{l\neq n};p,\{q_{l}\}\right)}{T_{2}\left(p+\sum_{l}q_{l}-\sum_{l\neq n}k_{l}\right)}=\sum_{j=1}^{n}A_{j}+\sum_{i=1}^{n-1}B_{i}+C,\label{eq:GammaNP1}
\end{equation}
where by default the index $l$ runs from $1$ to $n$, and \begin{widetext}
\begin{eqnarray}
A_{j} & = & \left(-1\right)^{j-1}G_{0\downarrow}\left(p+\sum_{l\neq n-j+1}q_{l}-\sum_{l\neq n}k_{l}\right)\Gamma_{n}\left(k_{1}\cdots k_{n-2};p,q_{1}\cdots q_{n-j}q_{n-j+2}\cdots q_{n}\right),\label{eq: GammaNP1Aj}\\
B_{i} & = & \sum_{K}G_{0\uparrow}\left(K\right)G_{0\downarrow}\left(p+\sum_{l}q_{l}-\sum_{l\neq n}k_{l}-K\right)\Gamma_{n+1}\left(k_{1}\cdots k_{n-i-1}Kk_{n-i+1}\cdots k_{n-1};p,q_{1}\cdots q_{n}\right),\label{eq: GammaNP1Bi}\\
C & = & -\sum_{KQ}G_{0\uparrow}\left(K\right)G_{0\downarrow}\left(p+\sum_{l}q_{l}-\sum_{l\neq n}k_{l}-K\right)G_{0\uparrow}\left(Q\right)\Gamma_{n+2}\left(k_{1}\cdots k_{n-1}K;p,q_{1}\cdots q_{n}Q\right),
\end{eqnarray}
\end{widetext}In $A_{j}$, the argument $q_{n+1-j}$ is absent in
$\Gamma_{n}$. As explained in detail in Appendix A, a sign factor
$(-1)^{j-1}$ arises due to the antisymmetrization among $\{q_{l}\}$.
In $B_{i}$, the argument $K$ of $\Gamma_{n+1}$ is located at the
position $n-i$. $B_{i}$ always has a positive sign. We acquire a
sign factor of $(-1)^{i-1}$ if we move the argument $K$ all the
way to the right-hand-side of $k_{n-1}$. The sign factor $(-1)^{i-1}$
then makes $\sum_{i=1}^{n-1}B_{i}$ antisymmetric, upon the exchange
of any two momenta among $\{k_{l}\}_{l\neq n}$.

Eq. (\ref{eq: Gamma2}) for $\Gamma_{2}(k;p,q)$ and Eq. (\ref{eq:GammaNP1})
for $\Gamma_{n+1}(\{k_{l}\}_{l\neq n};p,\{q_{l}\})$ form a complete
set of equations to determine all the vertex functions. This set of
equations encloses at a particular order $n$, if we discard the $C$-term
in $\Gamma_{n+1}(\{k_{l}\}_{l\neq n};p,\{q_{l}\})$. In particular,
if we neglect the last shaded diagrams in Fig. \ref{fig3: Gamma3}
or in Fig. \ref{fig6: Gamma4}, we have the enclosed set of equations
to calculate the vertex functions, with the inclusion of two-particle-hole
excitations or three-particle-hole excitations. 

\begin{figure}[b]
\begin{centering}
\includegraphics[width=0.5\textwidth]{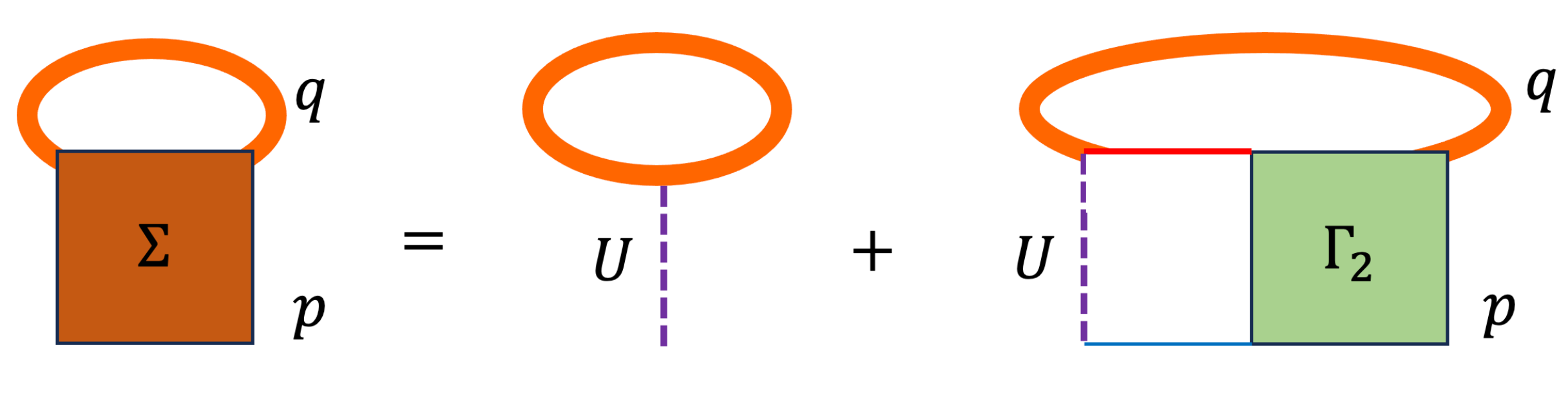}
\par\end{centering}
\caption{\label{fig7: DysonEquation} Dyson equation for the impurity, which
expresses the impurity self-energy $\Sigma(p)$ in terms of the vertex
function $\Gamma_{2}(k;p,q)$. The first Hartree diagram does not
contribute, due to the vanishingly small interaction strength $U\rightarrow0^{-}$.}
\end{figure}

\subsection{Dyson equation for the polaron self-energy}

As we are interested in obtaining the polaron spectral function, another
fundamental quantity is the polaron self-energy $\Sigma(p)$, whose
diagrammatic representation is given in Fig. \ref{fig7: DysonEquation},
according to the well-known Dyson equation \citep{TextbookAGD1975}.
As the interaction strength $U$ is infinitesimal, the first Hartree
diagram with a single dashed line does not contribute. For the second
diagram, we may write down,

\begin{equation}
\Sigma\left(p\right)=\sum_{q}G_{0\uparrow}\left(q\right)\gamma_{2}\left(p,q\right),\label{eq: SelfEnergy}
\end{equation}
where we have introduce a new variable 
\begin{equation}
\gamma_{2}\left(p,q\right)=-U\sum_{k}G_{0\uparrow}\left(k\right)G_{0\downarrow}\left(p+q-k\right)\Gamma_{2}\left(k;p,q\right).
\end{equation}
Therefore, we may directly calculate the self-energy $\Sigma(p)$,
once we obtain the vertex function $\Gamma_{2}(k;p,q)$. In turn,
we determine the impurity Green function ($p\equiv\{\mathbf{p},\omega\}$),
\begin{equation}
G_{\downarrow}\left(p\right)=\frac{1}{\omega-\varepsilon_{\mathbf{p}}^{I}-\Sigma(p)},
\end{equation}
and obtain the polaron spectral function $A(\mathbf{p},\omega)=-\textrm{Im}G_{\downarrow}(p)/\pi$
\citep{TextbookMahan2000,TextbookAGD1975}.

\subsection{On-shell vertex functions}

We now have an exact set of expressions, each of which involves the
summation over four-momenta. Here, we would like to point out that,
in the single-impurity or polaron limit, the summation over the fermionic
Matsubara frequency can be explicitly carried out \citep{Combescot2007,Hu2022b}.
As we shall see, we have the summation over the two kinds of four-momentum.
The one denoted by the variables $k$ (or $K$) is particle-like and
is the momentum of a fermion line propagating forward. For example,
we have the summation $\sum_{k}G_{0\uparrow}(k)P(k,\cdots)$, where
the function $P$ contains the particle-like momentum $k$ and other
variables in $\{\cdots\}$, and does not have any singularity at positive
energy. Another is hole-like and is the momentum of a Fermi loop moving
backward. It will be denoted by the variables $q$ (or $Q$), so we
have the summation such as $\sum_{q}G_{0\uparrow}(q)H(q,\cdots)$,
where the function $H$ is analytic on the left half-plane with negative
energy. 

As discussed in detail in Appendix B, we find two simple rules related
to the summation over the fermionic Matsubara frequency part of $k$
and $q$,
\begin{eqnarray}
\sum_{k}G_{0\uparrow}\left(k\right)P\left(k,\cdots\right) & = & -\sum_{\mathbf{k}}f\left(-\xi_{\mathbf{k}}\right)P\left(\{\mathbf{k},\xi_{\mathbf{k}}\},\cdots\right),\label{eq: MatFreqSumRule1}\\
\sum_{q}G_{0\uparrow}\left(q\right)H\left(q,\cdots\right) & = & \sum_{\mathbf{q}}f\left(\xi_{\mathbf{q}}\right)H\left(\{\mathbf{q},\xi_{\mathbf{q}}\},\cdots\right),\label{eq: MatFreqSumRule2}
\end{eqnarray}
where $f(x)\equiv1/(e^{x/k_{B}T}+1)$ is the Fermi-Dirac distribution
function, and $\xi_{\mathbf{k}}=\varepsilon_{\mathbf{k}}-\mu$ and
$\xi_{\mathbf{q}}=\varepsilon_{\mathbf{q}}-\mu$ are the dispersion
relations of fermions, measured from the chemical potential $\mu$.
In other words, the action of the summation over the fermionic Matsubara
frequency merely turns the varying Matsubara frequency into an on-shell
value, either $\xi_{\mathbf{k}}$ or $\xi_{\mathbf{q}}$, with a weighting
factor of $f(-\xi_{\mathbf{k}})$ or $f(\xi_{\mathbf{q}})$, which
is related to the thermal occupation of quasiparticle states, for
either particles or holes.

By applying these two rules for the summation over the fermionic Matsubara
frequency (i.e., for the particle-like momentum and hole-like momentum,
respectively), it is straightforward to obtain from Eq. (\ref{eq: Gamma2})
that,\begin{widetext}

\begin{equation}
\Gamma_{2}\left(k;p,q\right)=T_{2}\left(p+q\right)+\sum_{\mathbf{q}'}\Gamma_{3}\left(k;p,qq'\right)f\left(\xi_{\mathbf{q}'}\right)+\sum_{\mathbf{k}'\mathbf{q}'}T_{2}\left(p+q\right)G_{0\downarrow}\left(p+q-k'\right)\Gamma_{3}\left(k';p,qq'\right)f\left(-\xi_{\mathbf{k'}}\right)f\left(\xi_{\mathbf{q}'}\right),\label{eq: Gamma2R}
\end{equation}
where we have now used the notations $k'\equiv\{\mathbf{k'},\xi_{\mathbf{k}'}$\}
and $q'\equiv\{\mathbf{q'},\xi_{\mathbf{q'}}\}$. From this expression,
it becomes clear that we only need to take care of the vertex functions
with on-shell four-momenta. For example, in the vertex function $\Gamma_{2}(k;p,q)$,
we could take $k\equiv\{\mathbf{k},\xi_{\mathbf{k}}\}$ and $q\equiv\{\mathbf{q},\xi_{\mathbf{q}}\}$,
as the information of off-shell values is completely not needed. Hereafter,
we will always assume the on-shell values for any four-momentum. Of
course, the four-momentum $p=\{\mathbf{p},\omega\}$ for the impurity
is special. Here, the spatial momentum $\mathbf{p}$ and the energy
$\omega$ are the given input parameters, for the calculation of the
polaron spectral function. For the (on-shell) three-particle vertex
function, we then similarly obtain its expression,
\begin{eqnarray}
\frac{\Gamma_{3}\left(k;p,qq'\right)}{T_{2}\left(p+q+q'-k\right)} & = & +G_{0\downarrow}\left(p+q-k\right)\Gamma_{2}\left(k;p,q\right)-G_{0\downarrow}\left(p+q'-k\right)\Gamma_{2}\left(k;p,q'\right)\nonumber \\
 &  & -\sum_{\mathbf{k}''}G_{0\downarrow}\left(p+q+q'-k-k''\right)\left[\Gamma_{3}\left(k'';p,qq'\right)-\sum_{\mathbf{q}''}\Gamma_{4}\left(kk'';p,qq'q''\right)f\left(\xi_{\mathbf{q}''}\right)\right]f\left(-\xi_{\mathbf{k}''}\right).\label{eq: Gamma3R}
\end{eqnarray}
Finally, for the self-energy we obtain
\begin{equation}
\Sigma\left(p\right)=\sum_{\mathbf{q}}f\left(\xi_{\mathbf{q}}\right)\gamma_{2}\left(p,\{\mathbf{q},\xi_{\mathbf{q}}\}\right),\label{eq: SelfEnergyR}
\end{equation}
where
\begin{equation}
\gamma_{2}\left(p,q\right)=U\sum_{\mathbf{k}}G_{0\downarrow}\left(p+q-k\right)\Gamma_{2}\left(k;p,q\right)f\left(-\xi_{\mathbf{k}}\right).\label{eq: g2}
\end{equation}
By using Eq. (\ref{eq: Gamma2R}) to replace $\Gamma_{2}(k;p,q)$
in $\gamma_{2}(p,q)$, we find an alternative expression for $\gamma_{2}(p,q)$,
\begin{equation}
\gamma_{2}\left(p,q\right)=T_{2}\left(p+q\right)+\sum_{\mathbf{k}'\mathbf{q}'}T_{2}\left(p+q\right)G_{0\downarrow}\left(p+q-k'\right)\Gamma_{3}\left(k';p,qq'\right)f\left(-\xi_{\mathbf{k}'}\right)f\left(\xi_{\mathbf{q}'}\right),\label{eq: g2R}
\end{equation}
\end{widetext}which differs from $\Gamma_{2}(k;p,q)$ only by a term
$\sum_{\mathbf{q}'}\Gamma_{3}(k;p,qq')f(\xi_{\mathbf{q}'})$, i.e.,
the second term in Eq. (\ref{eq: Gamma2R}). To understand the difference,
we note that, to derive Eq. (\ref{eq: g2R}), we have used the identities,
\begin{equation}
U\sum_{\mathbf{k}}G_{0\downarrow}\left(p+q-k\right)f\left(-\xi_{\mathbf{k}}\right)=1
\end{equation}
 and 
\begin{equation}
U\sum_{\mathbf{k}}G_{0\downarrow}\left(p+q-k\right)\Gamma_{3}\left(k;p,qq'\right)f\left(-\xi_{\mathbf{k}}\right)=0.
\end{equation}
Both identities are related to infinitesimal running interaction strength
$U$. In the former identify, $G_{0\downarrow}(p+q-k)=1/[\omega+\xi_{\mathbf{q}}-\xi_{\mathbf{k}}-\varepsilon_{\mathbf{p}+\mathbf{q}-\mathbf{k}}^{I}]\sim2m_{r}/(\hbar^{2}\mathbf{k}^{2})$
at large momentum $\left|\mathbf{k}\right|$ and therefore $\sum_{\mathbf{k}}G_{0\downarrow}(p+q-k)$
diverges. This divergence is precisely compensated by the infinitesimal
$U$, according to our regularization recipe Eq. (\ref{eq: Ureg}).
In contrast, in the latter identity, $\Gamma_{3}(k;p,qq')$ should
decay fast enough at large momentum $\left|\mathbf{k}\right|$ and
the integral $\sum_{\mathbf{k}}G_{0\downarrow}(p+q-k)\Gamma_{3}(k;p,qq')f(-\xi_{\mathbf{k}})$
becomes finite. The smallness of the running interaction strength
$U$ then makes the term to vanish.

As a brief summary, we have obtained an exact set of equations to
determine the polaron spectral function. It consists of Eq. (\ref{eq: SelfEnergyR})
and Eq. (\ref{eq: g2R}) for the self-energy, Eq. (\ref{eq: Gamma2R})
and Eq. (\ref{eq: Gamma3R}) for the vertex function and the three-particle
vertex function, as well as Eq. (\ref{eq:GammaNP1}), which, after
we integrate out the fermionic Matsubara frequency, gives the on-shell
multi-particle vertex functions $\Gamma_{4}$. $\Gamma_{5}$ and so
on.

To close this subsection, it is worth noting that, the use of an infinitesimal
contact interaction strength $U$ plays an important role to determine
the complete sets of diagrams for the multi-particle vertex functions.
It enables us to replace the bare, running interaction strength $U$
everywhere by the many-body $T$-matrix $T_{2}$, as a Feynman diagram
with a finite number of the dashed interaction line must vanish. Without
this useful property, we will end up with infinitely many messy Feynman
diagrams, which we may hardly sum up, even at the first order of taking
only one-particle-hole excitations. 

\subsection{Two-particle-hole excitations}

In Eq. (\ref{eq: g2R}) for $\gamma_{2}(p,q)$, if we only keep the
first term $T_{2}(p+q)$, we are able to calculate the self-energy
Eq. (\ref{eq: SelfEnergyR}) with one-particle-hole excitations only
\citep{Combescot2007,Hu2022b}. To include two or more particle-hole
excitations, we need to keep the second term related to $\Gamma_{3}$
in Eq. (\ref{eq: g2R}) and it is necessarily to determine the three-particle
vertex function $\Gamma_{3}(k;p,qq')$ in Eq. (\ref{eq: Gamma3R}),
which in turn may require the knowledge of $\Gamma_{4}$. Actually,
the exact set of equations that we have just derived has a very nice
structure in hierarchy. For example, as we discussed earlier, in Eq.
(\ref{eq:GammaNP1}) for the multi-particle vertex function $\Gamma_{n+1}$,
if we discard the last $C$-term, then the set of equations encloses,
up to the approximation of including $n$-particle-hole excitations.
Here, we would like to present the detailed equations up to the level
of including two-particle-hole excitations, which gives the first
non-trivial vertex correction, beyond the many-body $T$-matrix theory
of Fermi polarons.

\begin{figure}[b]
\begin{centering}
\includegraphics[width=0.5\textwidth]{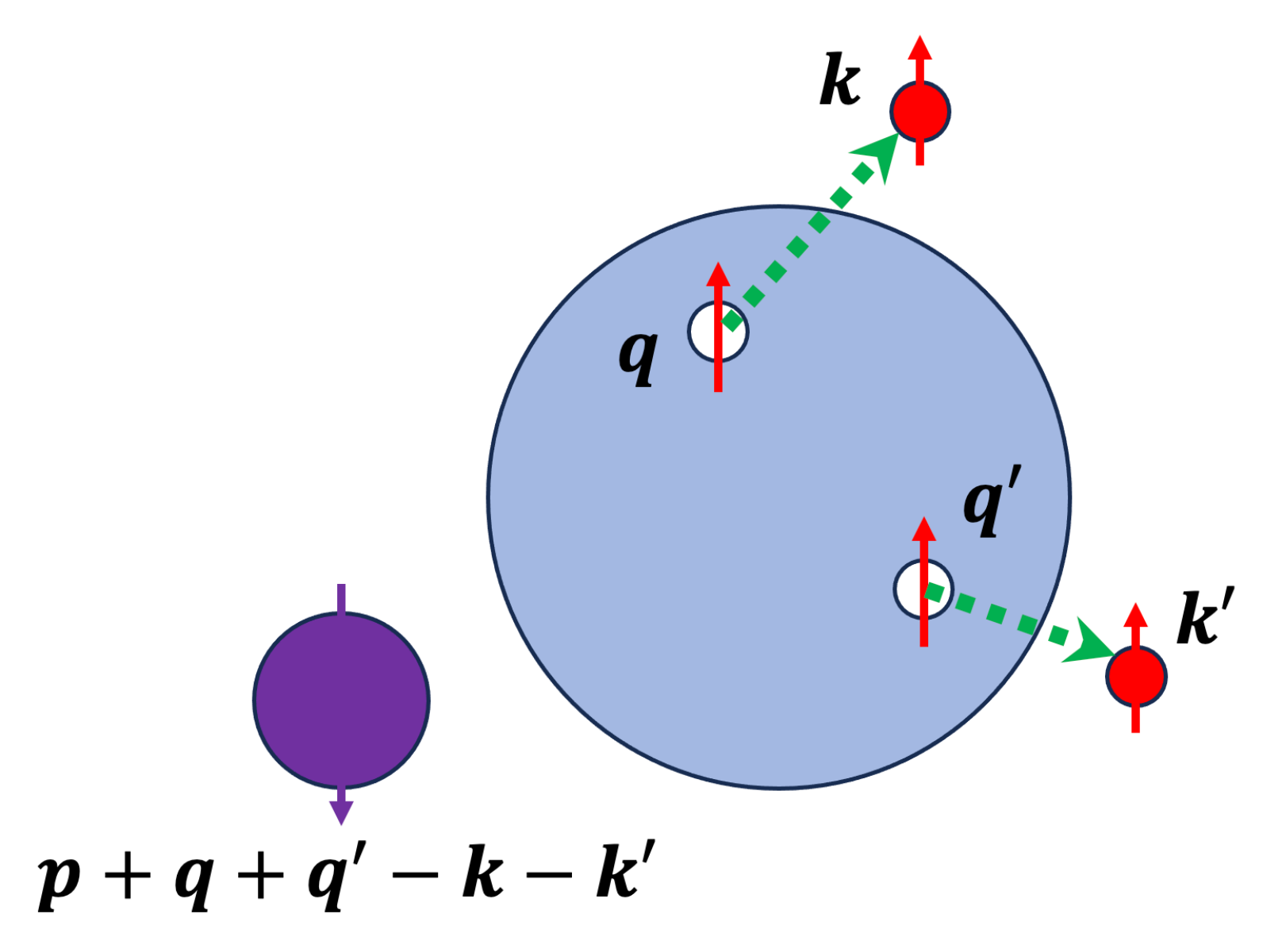}
\par\end{centering}
\caption{\label{fig8: FermiSea2ph} An illustration of a two-particle-hole
excitation out of the Fermi sea at zero temperature. The energy related
to such a two-particle-hole excitation is given by, $\bar{E}_{\mathbf{p};\mathbf{kk}';\mathbf{qq}'}^{(2)}=\varepsilon_{\mathbf{k}}+\varepsilon_{\mathbf{k}'}-\varepsilon_{\mathbf{q}}-\varepsilon_{\mathbf{q'}}+\varepsilon_{\mathbf{p}+\mathbf{q}+\mathbf{q'}-\mathbf{k}-\mathbf{k'}}^{I}-\varepsilon_{\mathbf{p}}^{I}$.
For an one-particle-hole excitation, we can similarly find the one-particle-hole
excitation energy $\bar{E}_{\mathbf{p};\mathbf{k};\mathbf{q}}^{(1)}=\varepsilon_{\mathbf{k}}-\varepsilon_{\mathbf{q}}+\varepsilon_{\mathbf{p}+\mathbf{q}-\mathbf{k}}^{I}-\varepsilon_{\mathbf{p}}^{I}$.}
\end{figure}

For this purpose, for a given four-momentum $p=\{\mathbf{p},\omega\}$,
it is useful to define the new variables,
\begin{eqnarray}
F_{\mathbf{q}} & \equiv & \gamma_{2}\left(p,q\right),\\
\alpha_{\mathbf{q}}^{\mathbf{k}} & \equiv & -\frac{\Gamma_{2}\left(k;p,q\right)}{E_{\mathbf{p};\mathbf{k};\mathbf{q}}^{(1)}},\\
G_{\mathbf{q}\mathbf{q}'}^{\mathbf{k}} & \equiv & \Gamma_{3}\left(k;p,qq'\right),
\end{eqnarray}
where $E_{\mathbf{p};\mathbf{k};\mathbf{q}}^{(1)}\equiv-G_{0\downarrow}^{-1}(p+q-k)=-\omega+\xi_{\mathbf{k}}-\xi_{\mathbf{q}}+\varepsilon_{\mathbf{p}+\mathbf{q}-\mathbf{k}}^{I}$
is basically the excitation energy of one-particle-hole excitations,
if we take $\omega=\varepsilon_{\mathbf{p}}^{I}$. The equation for
the three-particle vertex function Eq. (\ref{eq: Gamma3R}) then turns
into,

\begin{equation}
\frac{G_{\mathbf{q}\mathbf{q}'}^{\mathbf{k}}}{T_{2}\left(p+q+q'-k\right)}=\alpha_{\mathbf{q}}^{\mathbf{k}}-\alpha_{\mathbf{q}'}^{\mathbf{k}}+\sum_{\mathbf{k}''}\frac{G_{\mathbf{q}\mathbf{q}'}^{\mathbf{k}''}f\left(-\xi_{\mathbf{k}''}\right)}{E_{\mathbf{p};\mathbf{kk''};\mathbf{qq'}}^{(2)}},\label{eq: Gamma3R2}
\end{equation}
where $E_{\mathbf{p};\mathbf{kk''};\mathbf{qq'}}^{(2)}\equiv-G_{0\downarrow}^{-1}\left(p+q+q'-k-k''\right)=-\omega+\xi_{\mathbf{k}}+\xi_{\mathbf{k''}}-\xi_{\mathbf{q}}-\xi_{\mathbf{q}'}+\varepsilon_{\mathbf{p}+\mathbf{q}+\mathbf{q}'-\mathbf{k}-\mathbf{k}''}^{I}$
is the excitation energy of two-particle-hole excitations, if we again
take $\omega=\varepsilon_{\mathbf{p}}^{I}$; see, for example, Fig.
\ref{fig8: FermiSea2ph}. The equation for $\gamma_{2}(p,q)$, Eq.
(\ref{eq: g2R}), takes the form, 
\begin{equation}
F_{\mathbf{q}}=T_{2}\left(p+q\right)\left[1-\sum_{\mathbf{k}'\mathbf{q}'}\frac{G_{\mathbf{q}\mathbf{q}'}^{\mathbf{k}'}}{E_{\mathbf{p};\mathbf{k}';\mathbf{q}}^{(1)}}f\left(-\xi_{\mathbf{k}'}\right)f\left(\xi_{\mathbf{q}'}\right)\right].\label{eq: g2R2}
\end{equation}
In turn, the vertex function is now given by,
\begin{equation}
\Gamma_{2}\left(k;p,q\right)=F_{\mathbf{q}}+\sum_{\mathbf{q}'}G_{\mathbf{q}\mathbf{q}'}^{\mathbf{k}}f\left(\xi_{\mathbf{q}'}\right).\label{eq: Gamma2R2}
\end{equation}
Eq. (\ref{eq: Gamma3R2}), Eq. (\ref{eq: g2R2}), Eq. (\ref{eq: Gamma2R2}),
together with the definition of $\alpha_{\mathbf{q}}^{\mathbf{k}}$,
form a closed set of equations, which can be solved to determine the
self-energy $\Sigma(p)=\sum_{\mathbf{q}}f(\xi_{\mathbf{q}})F_{\mathbf{q}}$.
To make the equations complete, we also list the explicit expression
of the many-body $T$-matrix $T_{2}$ \citep{Hu2022b},
\begin{equation}
T_{2}\left(Q\equiv\{\mathbf{Q},\Omega\}\right)=\left[\frac{1}{U}-\sum_{\mathbf{K}}\frac{f\left(-\xi_{\mathbf{K}}\right)}{\Omega-\xi_{\mathbf{K}}-\varepsilon_{\mathbf{Q}-\mathbf{K}}^{I}}\right],\label{eq: T2}
\end{equation}
where the running interaction strength $U$ is to be replaced by the
$s$-wave scattering length $a$, for example, by using Eq. (\ref{eq: Ureg})
in three dimensions.

\section{Chevy ansatz }

Let us now turn to consider the Chevy ansatz approach \citep{Chevy2006,Combescot2008,Liu2022}.
Chevy ansatz has been widely used as a variational approach at zero
temperature. Its finite-temperature extension has also recently been
considered, with the inclusion of one-particle-hole excitations \citep{Liu2019}.
Here, we aim to generalize Chevy ansatz to finite temperature, including
arbitrary numbers of particle-hole excitations. We would like to comment
from the beginning that, at nonzero temperature, it probably does
not make sense to emphasize the variational aspect of Chevy ansatz,
since the polaron state is no longer a single quantum many-body state.
Therefore, even for the attractive polaron, its energy at a given
finite temperature may increase with the inclusion of more particle-hole
excitations. Actually, as we shall see, the most important feature
of Chevy ansatz is the closure of the Hilbert space for available
quantum states, if we truncate the ansatz to a particular level of
$n$-particle-hole excitations. This may also explain why we can find
out the complete set of diagrams for the multi-particle vertex functions,
as we discussed in the last section.

\subsection{Chevy ansatz at finite temperature}

Following the seminal works by Chevy \citep{Chevy2006} and Combescot
and Giraud \citep{Combescot2008,Giraud2010}, we take the following
Chevy ansatz for Fermi polarons (with momentum $\mathbf{p}$),\begin{widetext}
\begin{eqnarray}
\left|\psi\right\rangle  & = & \left[\alpha_{0}d_{\mathbf{p}}^{\dagger}+\sum_{\mathbf{kq}}\alpha_{\mathbf{q}}^{\mathbf{k}}d_{\mathbf{p}+\mathbf{q}-\mathbf{k}}^{\dagger}c_{\mathbf{k}}^{\dagger}c_{\mathbf{q}}+\frac{1}{2!2!}\sum_{\mathbf{kk'qq'}}\alpha_{\mathbf{q}\mathbf{q}'}^{\mathbf{k}\mathbf{k}'}d_{\mathbf{p}+\mathbf{q}+\mathbf{q}'-\mathbf{k}-\mathbf{k}'}^{\dagger}c_{\mathbf{k}}^{\dagger}c_{\mathbf{k}'}^{\dagger}c_{\mathbf{q'}}c_{\mathbf{q}}+\cdots\right]\left|\textrm{FS}\right\rangle ,\\
 & = & \sum_{n=0}^{\infty}\frac{1}{\left(n!\right)^{2}}\sum_{\mathbf{k}_{1}\cdots\mathbf{k}_{n}\mathbf{q}_{1}\cdots\mathbf{q}_{n}}\alpha_{\mathbf{q}_{1}\cdots\mathbf{q}_{n}}^{\mathbf{k}_{1}\cdots\mathbf{k}_{n}}d_{\mathbf{p}+(\mathbf{q}_{1}+\cdots+\mathbf{q}_{n})-(\mathbf{k}_{1}+\cdots+\mathbf{k}_{n})}^{\dagger}c_{\mathbf{k}_{1}}^{\dagger}\cdots c_{\mathbf{k}_{n}}^{\dagger}c_{\mathbf{q}_{n}}\cdots c_{\mathbf{q}_{1}}\left|\textrm{FS}\right\rangle \equiv\sum_{n=0}^{\infty}\left|\psi_{n}\right\rangle ,\label{eq: ChevyAnsatz}
\end{eqnarray}
\end{widetext}where $\left|\textrm{FS}\right\rangle $ describes
a thermal Fermi sea at finite temperature, in which the occupation
of a single-particle state with momentum $\mathbf{k}$ is given by
the Fermi-Dirac distribution function $f(\xi_{\mathbf{k}})$. At zero
temperature, we can find the sharp Fermi surface located at the Fermi
wavevector $k_{F}$ and clearly distinguish a particle excitation
with $\left|\mathbf{k}_{i}\right|>k_{F}$ from a hole excitation with
$\left|\mathbf{q}_{i}\right|<k_{F}$. At nonzero temperature, the
Fermi surface becomes blurred. Nevertheless, we will relax our definition
of particle or hole excitations. We may still use $\mathbf{k}_{i}$
to denote a ``particle''-like state out of the Fermi sea and $\mathbf{q}_{i}$
to denote a ``hole''-like state within the Fermi sea. It is also
convenient to use the abbreviations, $\{\mathbf{k}\}\equiv\{\mathbf{k}_{1}\cdots\mathbf{k}_{n}\}$,
$\{\mathbf{q}\}\equiv\{\mathbf{q}_{1}\cdots\mathbf{q}_{n}\}$, and
$\{\mathbf{kq}\}\equiv\{\mathbf{k}_{1}\cdots\mathbf{k}_{n}\mathbf{q}_{1}\cdots\mathbf{q}_{n}\}$.
In the case of potential confusion, we will take the full labelling. 

It is important to note that, the definition of the coefficients $\alpha_{\{\mathbf{q}\}}^{\{\mathbf{k}\}}$
in our ansatz Eq. (\ref{eq: ChevyAnsatz}) is slightly different from
what used in the earlier works \citep{Chevy2006,Combescot2008,Giraud2010,Liu2022}.
Here, we re-arrange the order of the annihilation field operators
$c_{\mathbf{q}}$ in the ansatz. For example, for the two-particle-hole
excitation term, we use $\cdots c_{\mathbf{q'}}c_{\mathbf{q}}\left|\textrm{FS}\right\rangle $,
instead of $\cdots c_{\mathbf{q}}c_{\mathbf{q'}}\left|\textrm{FS}\right\rangle $.
This re-arrangement is crucial to establish the relationship with
the diagrammatic theory in the last section. It will also remove some
unnecessary signs for the general expressions of the coupled equations
for the coefficients. We note also that, in the Chevy ansatz the coefficients
$\alpha_{\{\mathbf{q}\}}^{\{\mathbf{k}\}}$ should be antisymmetric
with respect to the exchange of $\mathbf{k}_{i}$ and $\mathbf{k}_{j}$
or the exchange of $\mathbf{q}_{i}$ and $\mathbf{q}_{j}$. As a consequence,
there are a lot of identical terms in Chevy ansatz, i.e., $\left(n!\right)^{2}$
terms, since the permutations either in $\{\mathbf{k}\}$ or in $\{\mathbf{q}\}$
generates a factorial $n!$. This redundancy can be simply removed
by the factor $1/(n!)^{2}$.

Our task now is to formally solve the stationary Schrödinger equation,
$\mathcal{H}\left|\psi\right\rangle =E\left|\psi\right\rangle $,
and derive the set of equations satisfied by the coefficients $\alpha_{\{\mathbf{q}\}}^{\{\mathbf{k}\}}$.
To proceed, it is useful to introduce some new variables. We will
use 
\begin{equation}
\left|\vec{\kappa}_{n}\right\rangle \equiv c_{\mathbf{k}_{1}}^{\dagger}\cdots c_{\mathbf{k}_{n}}^{\dagger}c_{\mathbf{q}_{n}}\cdots c_{\mathbf{q}_{1}}\left|\textrm{FS}\right\rangle 
\end{equation}
to describe the shake-up Fermi sea with a $n$-particle-hole excitation
out of $\left|\textrm{FS}\right\rangle $. It has a momentum $\mathbf{P}_{\vec{\kappa}_{n}}=(\mathbf{k}_{1}+\cdots+\mathbf{k}_{n})-(\mathbf{q}_{1}+\cdots+\mathbf{q}_{n})$
and an excitation energy $E_{\vec{\kappa}_{n}}=(\varepsilon_{\mathbf{k}_{1}}+\cdots+\varepsilon_{\mathbf{k}_{n}})-(\varepsilon_{\mathbf{q}_{1}}+\cdots+\varepsilon_{\mathbf{q}_{n}})$.
Thus, the $n$-th component of the Chevy ansatz, $\left|\psi_{n}\right\rangle $,
is given by,
\begin{equation}
\left|\psi_{n}\right\rangle =\frac{1}{\left(n!\right)^{2}}\sum_{\{\mathbf{kq}\}}\alpha_{\{\mathbf{q}\}}^{\{\mathbf{k}\}}d_{\mathbf{p}-\mathbf{P}_{\vec{\kappa}_{n}}}^{\dagger}\left|\vec{\kappa}_{n}\right\rangle .
\end{equation}
Let us check the results of $\mathcal{H}\left|\psi\right\rangle =(\mathcal{H}_{0}+\mathcal{H}_{\textrm{int}})\left|\psi\right\rangle $.

The action of the non-interacting, kinetic part of the Hamiltonian,
$\mathcal{H}_{0}\left|\psi\right\rangle $ (or specifically $\mathcal{H}_{0}\left|\psi_{n}\right\rangle $),
is straightforward to calculate. The creation field operator $d_{\mathbf{p}-\mathbf{P}_{\vec{\kappa}_{n}}}^{\dagger}$
creates an impurity with single-particle energy $\varepsilon_{\mathbf{p}-\mathbf{P}_{\vec{\kappa}_{n}}}^{I}$,
which is to be picked up by the term $\sum_{\mathbf{p}}\varepsilon_{\mathbf{p}}^{I}d_{\mathbf{p}}^{\dagger}d_{\mathbf{p}}$
in $\mathcal{H}_{0}$. In a similar way, the total single-particle
energy of $\left|\vec{\kappa}_{n}\right\rangle $ is $E_{\textrm{FS}}+E_{\vec{\kappa}_{n}}$,
where $E_{\textrm{FS}}$ is the energy of the unperturbed Fermi sea
$\left|\textrm{FS}\right\rangle $ at finite temperature. This total
single-particle energy will be picked up by the term $\sum_{\mathbf{k}}\varepsilon_{\mathbf{k}}c_{\mathbf{k}}^{\dagger}c_{\mathbf{k}}$
in $\mathcal{H}_{0}$. Putting together, we obtain,
\begin{equation}
\mathcal{H}_{0}\left|\psi_{n}\right\rangle =\sum_{\{\mathbf{kq}\}}\frac{\varepsilon_{\mathbf{p}-\mathbf{P}_{\vec{\kappa}_{n}}}^{I}+E_{\vec{\kappa}_{n}}+E_{\textrm{FS}}}{\left(n!\right)^{2}}\alpha_{\{\mathbf{q}\}}^{\{\mathbf{k}\}}d_{\mathbf{p}-\mathbf{P}_{\vec{\kappa}_{n}}}^{\dagger}\left|\vec{\kappa}_{n}\right\rangle .
\end{equation}
On the other hand, the action of the interacting Hamiltonian $\mathcal{H}_{\textrm{int}}=U\sum_{\mathbf{KK}'\mathbf{QQ}'}\delta_{\mathbf{K}+\mathbf{Q},\mathbf{K}'+\mathbf{Q}'}c_{\mathbf{K}}^{\dagger}c_{\mathbf{K}'}d_{\mathbf{Q}}^{\dagger}d_{\mathbf{Q}'}$
on the wave-function, $\mathcal{H}_{\textrm{int}}\left|\psi\right\rangle $,
is not so obvious to determine. As $\left|\psi_{n}\right\rangle $
involves only a single impurity with momentum $\mathbf{p}-\mathbf{P}_{\vec{\kappa}_{n}}$,
the action of the operators $d_{\mathbf{Q}}^{\dagger}d_{\mathbf{Q}'}$
simply changes the momentum of the impurity to $\mathbf{p}-\mathbf{P}_{\vec{\kappa}_{n}}+\mathbf{Q}-\mathbf{Q}'$.
In contrast, as the perturbed Fermi sea state $\left|\vec{\kappa}_{n}\right\rangle $
involves infinitely many spin-up fermions, the action of $c_{\mathbf{K}}^{\dagger}c_{\mathbf{K}'}$
on $\left|\vec{\kappa}_{n}\right\rangle $ should be analyzed carefully.
In addition to an overall Hartree shift $\nu U$, where $\nu$ is
the density of spin-up fermions, generally we find three different
situations, referred to as the $A$-, $B$-, and $C$-cases, if we
analyze the sector of the $n$-particle-hole excitations for $\mathcal{H}_{\textrm{int}}\left|\psi\right\rangle $.

In the $A$-case, we consider the action of $c_{\mathbf{K}}^{\dagger}c_{\mathbf{K}'}$
on $\left|\vec{\kappa}_{n-1}\right\rangle $, which creates a new
particle-hole excitation (where $\mathbf{K}$ is the particle momentum
and $\mathbf{K}'$ is the hole momentum), on top of the existing $(n-1)$-particle-hole
excitation. The resulting state is $\left|\vec{\kappa}'_{n}\right\rangle $,
where the prime indicate a new set of particle or hole momenta.

In the $B$-case, we take the action of $c_{\mathbf{K}}^{\dagger}c_{\mathbf{K}'}$
on $\left|\vec{\kappa}_{n}\right\rangle $, which does not change
the number of particle-hole excitations (i.e., we will end up with
$\left|\vec{\kappa}''_{n}\right\rangle $). Therefore, it is necessarily
to destroy an existing particle state or hole state in $\left|\vec{\kappa}_{n}\right\rangle $.
We also need to trace the momentum of this particle state or hole
state, over the existing momenta $\{\mathbf{k}\}$ or $\{\mathbf{q}\}$.

In the $C$-case, we apply the operators $c_{\mathbf{K}}^{\dagger}c_{\mathbf{K}'}$
on $\left|\vec{\kappa}_{n+1}\right\rangle $. In order to bring the
state back to $\left|\vec{\kappa}_{n}\right\rangle $ as we wish to
focus on the sector of $n$-particle-hole excitations, we need to
destroy both a particle state (with the momentum $\mathbf{K}'$) and
a hole state (with the momentum $\mathbf{K}$) in $\left|\vec{\kappa}_{n+1}\right\rangle $.

By enumerating all the possibilities, after a lengthy algebra, we
obtain ($\mathcal{\bar{H}}_{\textrm{int}}\equiv\mathcal{H}_{\textrm{int}}-\nu U$),
\begin{equation}
\mathcal{\bar{H}}_{\textrm{int}}\left|\psi\right\rangle =\sum_{n=0}^{\infty}\sum_{\{\mathbf{kq}\}}\frac{A_{\{\mathbf{q}\}}^{\{\mathbf{k}\}}+B_{\{\mathbf{q}\}}^{\{\mathbf{k}\}}+C_{\{\mathbf{q}\}}^{\{\mathbf{k}\}}}{\left(n!\right)^{2}}d_{\mathbf{p}-\mathbf{P}_{\vec{\kappa}_{n}}}^{\dagger}\left|\vec{\kappa}_{n}\right\rangle ,
\end{equation}
where
\begin{eqnarray}
A_{\{\mathbf{q}\}}^{\{\mathbf{k}\}} & = & U\sum_{i,j=1,\cdots,n}\left(-1\right)^{i+j}\alpha_{\mathbf{q}_{1}\cdots\mathbf{q}_{n-j}\mathbf{q}_{n-j+2}\cdots\mathbf{q}_{n}}^{\mathbf{k}_{1}\cdots\mathbf{k}_{n-i}\mathbf{k}_{n-i+2}\cdots\mathbf{k}_{n}},\\
B_{\{\mathbf{q}\}}^{\{\mathbf{k}\}} & = & U\sum_{\mathbf{K}}\left(\alpha_{\mathbf{q}_{1}\cdots\mathbf{q}_{n}}^{\mathbf{K}\cdots\mathbf{k}_{n}}+\cdots+\alpha_{\mathbf{q}_{1}\cdots\mathbf{q}_{n}}^{\mathbf{k}_{1}\cdots\mathbf{K}}\right)f\left(-\xi_{\mathbf{K}}\right)-\nonumber \\
 &  & U\sum_{\mathbf{Q}}\left(\alpha_{\mathbf{Q}\cdots\mathbf{q}_{n}}^{\mathbf{k}_{1}\cdots\mathbf{k}_{n}}+\cdots+\alpha_{\mathbf{q}_{1}\cdots\mathbf{Q}}^{\mathbf{k}_{1}\cdots\mathbf{k}_{n}}\right)f\left(\xi_{\mathbf{Q}}\right),\\
C_{\{\mathbf{q}\}}^{\{\mathbf{k}\}} & = & U\sum_{\mathbf{KQ}}\alpha_{\mathbf{q}_{1}\cdots\mathbf{q}_{n}\mathbf{Q}}^{\mathbf{k}_{1}\cdots\mathbf{k}_{n}\mathbf{K}}f\left(-\xi_{\mathbf{K}}\right)f\left(\xi_{\mathbf{Q}}\right),
\end{eqnarray}
All the three coefficients $A_{\{\mathbf{q}\}}^{\{\mathbf{k}\}}$,
$B_{\{\mathbf{q}\}}^{\{\mathbf{k}\}}$, $C_{\{\mathbf{q}\}}^{\{\mathbf{k}\}}$
are manifestly antisymmetric with respect to the exchange of any two
momenta in $\{\mathbf{k}\}$ or $\{\mathbf{q}\}$. The two Fermi distribution
functions $f\left(-\xi_{\mathbf{K}}\right)$ and $f\left(\xi_{\mathbf{Q}}\right)$
represent the availability of the particle state (with the momentum
$\mathbf{K}$) and of the hole state (with the momentum $\mathbf{Q}$),
respectively. In the expression of $B_{\{\mathbf{q}\}}^{\{\mathbf{k}\}}$,
a minus sign arises, simply because, to destroy a hole state we need
to rewrite $\cdots c_{\mathbf{K}}^{\dagger}c_{\mathbf{K}'}\cdots=-\cdots c_{\mathbf{K}'}c_{\mathbf{K}}^{\dagger}\cdots$
in the interaction Hamiltonian.

We are now ready to translate the stationary Schrödinger equation,
$(\mathcal{H-}E)\left|\psi\right\rangle =0$, into a set of equations
for the coefficients $\alpha_{\{\mathbf{q}\}}^{\{\mathbf{k}\}}$.
For this purpose, let use define the excitation energy,
\begin{equation}
E_{\mathbf{p};\{\mathbf{k}\};\{\mathbf{q}\}}^{(n)}\equiv\left(-E+E_{\textrm{FS}}+\nu U\right)+\varepsilon_{\mathbf{p}-\mathbf{P}_{\vec{\kappa}_{n}}}^{I}+E_{\vec{\kappa}_{n}},
\end{equation}
In the absence of the interaction $U=0$, $E_{\mathbf{p};\{\mathbf{k}\};\{\mathbf{q}\}}^{(n)}$
is precisely the energy cost for the creation of a $n$-particle-hole
excitation out of the Fermi sea, if we take $E=\varepsilon_{\mathbf{p}}^{I}+E_{\textrm{FS}}$.
In terms of $E_{\mathbf{p};\{\mathbf{k}\};\{\mathbf{q}\}}^{(n)}$,
we find that the coefficients $\alpha_{\{\mathbf{q}\}}^{\{\mathbf{k}\}}$
satisfy a series of coupled equations, \begin{widetext}

\begin{eqnarray}
-E_{\mathbf{p};\{\mathbf{k}\};\{\mathbf{q}\}}^{(n)}\alpha_{\mathbf{q}_{1}\cdots\mathbf{q}_{n}}^{\mathbf{k}_{1}\cdots\mathbf{k}_{n}} & = & U\sum_{i,j=1,\cdots,n}\left(-1\right)^{i+j}\alpha_{\mathbf{q}_{1}\cdots\mathbf{q}_{n-j}\mathbf{q}_{n-j+2}\cdots\mathbf{q}_{n}}^{\mathbf{k}_{1}\cdots\mathbf{k}_{n-i}\mathbf{k}_{n-i+2}\cdots\mathbf{k}_{n}}+U\sum_{\mathbf{K}}\left(\alpha_{\mathbf{q}_{1}\mathbf{q}_{2}\cdots\mathbf{q}_{n}}^{\mathbf{K}\mathbf{k}_{2}\cdots\mathbf{k}_{n}}+\cdots+\alpha_{\mathbf{q}_{1}\mathbf{q}_{2}\cdots\mathbf{q}_{n}}^{\mathbf{k}_{1}\cdots\mathbf{k}_{n-1}\mathbf{K}}\right)f\left(-\xi_{\mathbf{K}}\right)\nonumber \\
 &  & -U\sum_{\mathbf{Q}}\left(\alpha_{\mathbf{Q}\mathbf{q}_{2}\cdots\mathbf{q}_{n}}^{\mathbf{k}_{1}\mathbf{k}_{2}\cdots\mathbf{k}_{n}}+\cdots+\alpha_{\mathbf{q}_{1}\cdots\mathbf{q}_{n-1}\mathbf{Q}}^{\mathbf{k}_{1}\mathbf{k}_{2}\cdots\mathbf{k}_{n}}\right)f\left(\xi_{\mathbf{Q}}\right)+U\sum_{\mathbf{KQ}}\alpha_{\mathbf{q}_{1}\mathbf{q}_{2}\cdots\mathbf{q}_{n}\mathbf{Q}}^{\mathbf{k}_{1}\mathbf{k}_{2}\cdots\mathbf{k}_{n}\mathbf{K}}f\left(-\xi_{\mathbf{K}}\right)f\left(\xi_{\mathbf{Q}}\right).\label{eq:ChevyAnsatzSolution}
\end{eqnarray}
These equations connect the $n$-th coefficients $\alpha_{\{\mathbf{q}\}}^{\{\mathbf{k}\}}$
to the lower order $(n-1)$-th coefficients (see, i.e., the first
term in the right-hand-side of the equation) and the higher order
$(n+1)$-th coefficients (see the last term). The coupled equations
will enclose, if we neglect the last term, yielding the results valid
for the inclusion of $n$-particle-hole excitations.

\subsection{Chevy ansatz with two-particle-hole excitations}

To illustrate the usefulness of the exact set of equations Eq. (\ref{eq:ChevyAnsatzSolution}),
let us truncate to the order of $n=2$, as an example,
\begin{eqnarray}
\left(\tilde{E}-\varepsilon_{\mathbf{p}}^{I}\right)\alpha_{0} & = & U\sum_{\mathbf{KQ}}\alpha_{\mathbf{Q}}^{\mathbf{K}}f\left(-\xi_{\mathbf{K}}\right)f\left(\xi_{\mathbf{Q}}\right),\label{eq:CA2ph_E1}\\
-E_{\mathbf{p};\mathbf{k};\mathbf{q}}^{(1)}\alpha_{\mathbf{q}}^{\mathbf{k}} & = & U\alpha_{0}+U\sum_{\mathbf{K}}\alpha_{\mathbf{q}}^{\mathbf{K}}f\left(-\xi_{\mathbf{K}}\right)-U\sum_{\mathbf{Q}}\alpha_{\mathbf{Q}}^{\mathbf{k}}f\left(\xi_{\mathbf{Q}}\right)+U\sum_{\mathbf{KQ}}\alpha_{\mathbf{qQ}}^{\mathbf{kK}}f\left(-\xi_{\mathbf{K}}\right)f\left(\xi_{\mathbf{Q}}\right),\label{eq:CA2ph_E2}\\
-E_{\mathbf{p};\mathbf{kk'};\mathbf{qq'}}^{(2)}\alpha_{\mathbf{qq'}}^{\mathbf{kk'}} & = & U\left(\alpha_{\mathbf{q}}^{\mathbf{k}}-\alpha_{\mathbf{q}}^{\mathbf{k'}}-\alpha_{\mathbf{q'}}^{\mathbf{k}}+\alpha_{\mathbf{q'}}^{\mathbf{k}'}\right)+U\sum_{\mathbf{K}}\left(\alpha_{\mathbf{q}\mathbf{q}'}^{\mathbf{Kk'}}+\alpha_{\mathbf{q}\mathbf{q}'}^{\mathbf{kK}}\right)f\left(-\xi_{\mathbf{K}}\right)-U\sum_{\mathbf{Q}}\left(\alpha_{\mathbf{Q}\mathbf{q}'}^{\mathbf{kk'}}+\alpha_{\mathbf{qQ}}^{\mathbf{kk'}}\right)f\left(\xi_{\mathbf{Q}}\right).\label{eq:CA2ph_E3}
\end{eqnarray}
\end{widetext}Here we define $\tilde{E}=E-E_{\textrm{FS}}-\nu U$
as the energy of the Fermi polaron with the exclusion of the mean-field
Hartree shift $\nu U$. The explicit expressions of $E_{\mathbf{p};\mathbf{k};\mathbf{q}}^{(1)}$
and $E_{\mathbf{p};\mathbf{kk'};\mathbf{qq'}}^{(2)}$ are given by,
$E_{\mathbf{p};\mathbf{k};\mathbf{q}}^{(1)}=-\tilde{E}+\varepsilon_{\mathbf{p}+\mathbf{q}-\mathbf{k}}^{I}+\varepsilon_{\mathbf{k}}-\varepsilon_{\mathbf{q}}$
and $E_{\mathbf{p};\mathbf{kk'};\mathbf{qq'}}^{(2)}=-\tilde{E}+\varepsilon_{\mathbf{p}+\mathbf{q}+\mathbf{q}'-\mathbf{k}-\mathbf{k}'}^{I}+\varepsilon_{\mathbf{k}}+\varepsilon_{\mathbf{k'}}-\varepsilon_{\mathbf{q}}-\varepsilon_{\mathbf{q'}}$,
respectively. 

At zero temperature, where the Fermi distribution functions $f(-\xi_{\mathbf{K}})$
and $f(\xi_{\mathbf{Q}})$ respectively restrict the momentum $\left|\mathbf{K}\right|>k_{F}$
and $\left|\mathbf{Q}\right|<k_{F}$, the above three equations have
already been used to determine the attractive Fermi polaron energy
in free space, from one dimension to three dimensions \citep{Combescot2008,Parish2013}.
At nonzero temperature, we may discretize the momentum in the equations
to obtain the discrete polaron energy levels. However, it will be
more interesting to work with continuous momentum. In this case, we
may replace the energy $\tilde{E}$ by a frequency $\omega$. We may
take $\alpha_{0}=1$, if we do not care about the normalization of
the Chevy ansatz. As we shall see, the term $U\sum_{\mathbf{KQ}}\alpha_{\mathbf{Q}}^{\mathbf{K}}f(-\xi_{\mathbf{K}})f(\xi_{\mathbf{Q}})$
can then be understood as the polaron self-energy $\Sigma(\mathbf{p},\omega)$
at the given frequency $\omega$. 

To formally solve the coupled equations (at a given frequency $\omega$
and a given momentum $\mathbf{p}$), it is useful to define several
variables,
\begin{eqnarray}
F\left(\mathbf{q}\right) & \equiv & U+U\sum_{\mathbf{K}}\alpha_{\mathbf{q}}^{\mathbf{K}}f\left(-\xi_{\mathbf{K}}\right),\\
R\left(\mathbf{k}\right) & \equiv & U\sum_{\mathbf{Q}}\alpha_{\mathbf{Q}}^{\mathbf{k}}f\left(\xi_{\mathbf{Q}}\right),\\
G\left(\mathbf{k};\mathbf{q},\mathbf{q}'\right) & \equiv & U\sum_{\mathbf{K}}\alpha_{\mathbf{q}\mathbf{q}'}^{\mathbf{kK}}f\left(-\xi_{\mathbf{K}}\right),\\
H\left(\mathbf{k},\mathbf{k}';\mathbf{q}\right) & \equiv & U\sum_{\mathbf{Q}}\alpha_{\mathbf{qQ}}^{\mathbf{kk'}}f\left(\xi_{\mathbf{Q}}\right),\\
J\left(\mathbf{k},\mathbf{q}\right) & \equiv & \sum_{\mathbf{q}'}G\left(\mathbf{k};\mathbf{q},\mathbf{q}'\right)f\left(\xi_{\mathbf{q}'}\right).
\end{eqnarray}
For convenience, we often use the short-hand notations, $G_{\mathbf{qq}'}^{\mathbf{k}}\equiv G(\mathbf{k};\mathbf{q},\mathbf{q}')$
and $H_{\mathbf{q}}^{\mathbf{kk}'}\equiv H\left(\mathbf{k};\mathbf{k'};\mathbf{q}\right)$.
One may wonder that the short-hand notation $G_{\mathbf{qq}'}^{\mathbf{k}}$
has already used for the three-particle vertex function $\Gamma_{3}\left(k;p,qq'\right)$
in the diagrammatic theory. We will soon see that there is no contradiction,
as they are essentially the same quantity. According to these definitions,
we must satisfy the consistency conditions,
\begin{equation}
\Sigma\left(p\right)=\sum_{\mathbf{q}}\left[F\left(\mathbf{q}\right)-U\right]f\left(\xi_{\mathbf{q}}\right)=\sum_{\mathbf{k}}R\left(\mathbf{k}\right)f\left(-\xi_{\mathbf{k}}\right)
\end{equation}
and
\begin{equation}
J\left(\mathbf{k},\mathbf{q}\right)\equiv\sum_{\mathbf{q}'}G_{\mathbf{qq}'}^{\mathbf{k}}f\left(\xi_{\mathbf{q}'}\right)=\sum_{\mathbf{k}'}H_{\mathbf{q}}^{\mathbf{kk}'}f\left(-\xi_{\mathbf{k}'}\right).\label{eq: Jkq}
\end{equation}
The first equation Eq (\ref{eq:CA2ph_E1}) then turns into ($\alpha_{0}=1$
and $\nu=\sum_{\mathbf{q}}f(\xi_{\mathbf{q}})$),
\begin{equation}
\omega-\varepsilon_{\mathbf{p}}^{I}=\sum_{\mathbf{q}}F\left(\mathbf{q}\right)f\left(\xi_{\mathbf{q}}\right)-\nu U=\Sigma\left(\mathbf{p},\omega\right),\label{eq:CA2ph_SelfEnergy}
\end{equation}
which determines the pole of the polaron Green function. The second
equation Eq. (\ref{eq:CA2ph_E2}) becomes,
\begin{equation}
\frac{1}{U}\left(U\alpha_{\mathbf{q}}^{\mathbf{k}}\right)=-\frac{F\left(\mathbf{q}\right)-R\left(\mathbf{k}\right)+J\left(\mathbf{k},\mathbf{q}\right)}{E_{\mathbf{p};\mathbf{k};\mathbf{q}}^{(1)}}.\label{eq: akq}
\end{equation}
Let us multiply $f(-\xi_{\mathbf{k}})$ on both sides and integrate
over $\mathbf{k}$, or multiply $f(\xi_{\mathbf{q}})$ on both sides
and integrate over $\mathbf{q}$. These operations lead to two coupled
equations,\begin{widetext}
\begin{eqnarray}
F\left(\mathbf{q}\right) & = & \left[\frac{1}{U}+\sum_{\mathbf{k}}\frac{f\left(-\xi_{\mathbf{k}}\right)}{E_{\mathbf{p};\mathbf{k};\mathbf{q}}^{(1)}}\right]^{-1}\left[1+\sum_{\mathbf{k}}\frac{R\left(\mathbf{k}\right)-J\left(\mathbf{k},\mathbf{q}\right)}{E_{\mathbf{p};\mathbf{k};\mathbf{q}}^{(1)}}f\left(-\xi_{\mathbf{k}}\right)\right],\label{eq:CA2ph_Fq}\\
R\left(\mathbf{k}\right) & = & \left[\frac{1}{U}-\sum_{\mathbf{q}}\frac{f\left(\xi_{\mathbf{q}}\right)}{E_{\mathbf{p};\mathbf{k};\mathbf{q}}^{(1)}}\right]^{-1}\left[-\sum_{\mathbf{q}}\frac{F\left(\mathbf{q}\right)+J\left(\mathbf{k},\mathbf{q}\right)}{E_{\mathbf{p};\mathbf{k};\mathbf{q}}^{(1)}}f\left(\xi_{\mathbf{q}}\right)\right].\label{eq:CA2ph_Rk}
\end{eqnarray}
For the third equation Eq. (\ref{eq:CA2ph_E3}), we similarly find,
\begin{equation}
\frac{1}{U}\left(U\alpha_{\mathbf{qq'}}^{\mathbf{kk'}}\right)=-\frac{U\left(\alpha_{\mathbf{q}}^{\mathbf{k}}-\alpha_{\mathbf{q}}^{\mathbf{k'}}-\alpha_{\mathbf{q'}}^{\mathbf{k}}+\alpha_{\mathbf{q'}}^{\mathbf{k}'}\right)+\left(G_{\mathbf{qq}'}^{\mathbf{k}}-G_{\mathbf{qq}'}^{\mathbf{k'}}\right)-\left(H_{\mathbf{q}}^{\mathbf{kk}'}-H_{\mathbf{q'}}^{\mathbf{kk}'}\right)}{E_{\mathbf{p};\mathbf{kk'};\mathbf{qq'}}^{(2)}}.
\end{equation}
The integration over $\mathbf{k'}$ or $\mathbf{q}'$ then leads to
the coupled equations, 
\begin{eqnarray}
G_{\mathbf{qq}'}^{\mathbf{k}} & = & \left[\frac{1}{U}+\sum_{\mathbf{k'}}\frac{f\left(-\xi_{\mathbf{k'}}\right)}{E_{\mathbf{p};\mathbf{kk'};\mathbf{qq'}}^{(2)}}\right]^{-1}\left[-U\sum_{\mathbf{k'}}\frac{\alpha_{\mathbf{q}}^{\mathbf{k}}-\alpha_{\mathbf{q}}^{\mathbf{k'}}-\alpha_{\mathbf{q'}}^{\mathbf{k}}+\alpha_{\mathbf{q'}}^{\mathbf{k}'}}{E_{\mathbf{p};\mathbf{kk'};\mathbf{qq'}}^{(2)}}f\left(-\xi_{\mathbf{k'}}\right)+\sum_{\mathbf{k'}}\frac{G_{\mathbf{qq}'}^{\mathbf{k'}}+H_{\mathbf{q}}^{\mathbf{kk}'}-H_{\mathbf{q'}}^{\mathbf{kk}'}}{E_{\mathbf{p};\mathbf{kk'};\mathbf{qq'}}^{(2)}}f\left(-\xi_{\mathbf{k'}}\right)\right],\label{eq:CA2ph_Gkqqp}\\
H_{\mathbf{q}}^{\mathbf{kk}'} & = & \left[\frac{1}{U}-\sum_{\mathbf{q'}}\frac{f\left(\xi_{\mathbf{q'}}\right)}{E_{\mathbf{p};\mathbf{kk'};\mathbf{qq'}}^{(2)}}\right]^{-1}\left[-U\sum_{\mathbf{q'}}\frac{\alpha_{\mathbf{q}}^{\mathbf{k}}-\alpha_{\mathbf{q}}^{\mathbf{k'}}-\alpha_{\mathbf{q'}}^{\mathbf{k}}+\alpha_{\mathbf{q'}}^{\mathbf{k}'}}{E_{\mathbf{p};\mathbf{kk'};\mathbf{qq'}}^{(2)}}f\left(\xi_{\mathbf{q'}}\right)-\sum_{\mathbf{q'}}\frac{G_{\mathbf{qq}'}^{\mathbf{k}}-G_{\mathbf{qq}'}^{\mathbf{k'}}+H_{\mathbf{q'}}^{\mathbf{kk}'}}{E_{\mathbf{p};\mathbf{kk'};\mathbf{qq'}}^{(2)}}f\left(\xi_{\mathbf{q'}}\right)\right].\label{eq:CA2ph_Hkkpq}
\end{eqnarray}
\end{widetext}Eq. (\ref{eq:CA2ph_Fq}), Eq. (\ref{eq:CA2ph_Rk}),
Eq. (\ref{eq:CA2ph_Gkqqp}), and Eq. (\ref{eq:CA2ph_Hkkpq}) form
a closed set of coupled equations, together with the expressions of
$J(\mathbf{k},\mathbf{q})$ and $\alpha_{\mathbf{q}}^{\mathbf{k}}$,
Eq. (\ref{eq: Jkq}) and Eq. (\ref{eq: akq}). They might be numerically
solved in a self-consistent way, for any values of the interaction
strength $U$. However, to make the connection with the diagrammatic
theory, let us first check the case $U\rightarrow0^{-}$.

\subsection{Chevy ansatz at $U\rightarrow0^{-}$}

As we mentioned earlier, in two-dimensional or three-dimensional free
space, the contact interaction potential needs regularization, since
the integration over the high momentum diverges. Thus, we effectively
have a vanishingly small interaction strength $U\rightarrow0^{-}$.
In this case, the coupled equations of the Chevy ansatz solution Eq.
(\ref{eq:ChevyAnsatzSolution}) could be simplified. 

To find the rules of simplification, let us check the case of two-particle-hole
excitations. In the expressions for $R(\mathbf{k})$ and $H(\mathbf{k},\mathbf{k}';\mathbf{q})$,
see Eq. (\ref{eq:CA2ph_Rk}) and Eq. (\ref{eq:CA2ph_Hkkpq}), since
the integration over $\mathbf{q}$ (or $\mathbf{q}'$) is finite,
we have $[1/U-\sum_{\mathbf{q}}f(\xi_{\mathbf{q}})/E_{\mathbf{p};\mathbf{k};\mathbf{q}}^{(1)}]^{-1}\sim U$
and $[1/U-\sum_{\mathbf{q'}}f(\xi_{\mathbf{q'}})/E_{\mathbf{p};\mathbf{kk'};\mathbf{qq'}}^{(2)}]^{-1}\sim U$,
which are both infinitesimally small. Therefore, $R(\mathbf{k})=0$
and $H(\mathbf{k},\mathbf{k}';\mathbf{q})=0$. In general, in Eq.
(\ref{eq:ChevyAnsatzSolution}), the third term on the right-hand-side
of the equation vanishes identically. Moreover, in the expression
for $G(\mathbf{k};\mathbf{q},\mathbf{q}')$, Eq. (\ref{eq:CA2ph_Gkqqp}),
it is easy to see that,
\begin{equation}
-U\sum_{\mathbf{k'}}\frac{\left(-\alpha_{\mathbf{q}}^{\mathbf{k'}}+\alpha_{\mathbf{q'}}^{\mathbf{k}'}\right)f\left(-\xi_{\mathbf{k'}}\right)}{E_{\mathbf{p};\mathbf{kk'};\mathbf{qq'}}^{(2)}}\rightarrow0,
\end{equation}
since the integration over $\mathbf{k}'$ converges due to the well-behaved
coefficients $\alpha_{\mathbf{q}}^{\mathbf{k'}}\rightarrow0$ and
$\alpha_{\mathbf{q'}}^{\mathbf{k'}}\rightarrow0$ at large momentum
$\mathbf{k}'$. Also, we would have,
\begin{equation}
-\sum_{\mathbf{k'}}\frac{U\left(\alpha_{\mathbf{q}}^{\mathbf{k}}-\alpha_{\mathbf{q'}}^{\mathbf{k}}\right)f\left(-\xi_{\mathbf{k'}}\right)}{E_{\mathbf{p};\mathbf{kk'};\mathbf{qq'}}^{(2)}}=\alpha_{\mathbf{q}}^{\mathbf{k}}-\alpha_{\mathbf{q'}}^{\mathbf{k}},
\end{equation}
due to our regularization relation Eq. (\ref{eq: Ureg}) (i.e., the
smallness of $U$ exactly cancels the divergence of the integral over
$\mathbf{k}'$). More generally, therefore, only a few sub-terms in
the first term of Eq. (\ref{eq:ChevyAnsatzSolution}) contribute.
In fact, there are $n^{2}$ sub-terms in total. However, only $n$
sub-terms contribute. Following these observations, for $n\geq2$,
let us define the variables,
\begin{equation}
G_{\mathbf{q}_{1}\mathbf{q}_{2}\cdots\mathbf{q}_{n}}^{\mathbf{k}_{1}\mathbf{k}_{2}\cdots\mathbf{k}_{n-1}}\equiv U\sum_{\mathbf{K}}\alpha_{\mathbf{q}_{1}\mathbf{q}_{2}\cdots\mathbf{q}_{n}}^{\mathbf{k}_{1}\cdots\mathbf{k}_{n-1}\mathbf{K}}f\left(-\xi_{\mathbf{K}}\right).
\end{equation}
It is then straightforward to derive the equation for $G_{\mathbf{q}_{1}\mathbf{q}_{2}\cdots\mathbf{q}_{n}}^{\mathbf{k}_{1}\mathbf{k}_{2}\cdots\mathbf{k}_{n-1}}$:\begin{widetext}
\begin{eqnarray}
G_{\mathbf{q}_{1}\mathbf{q}_{2}\cdots\mathbf{q}_{n}}^{\mathbf{k}_{1}\mathbf{k}_{2}\cdots\mathbf{k}_{n-1}} & = & \left[\frac{1}{U}+\sum_{\mathbf{K}}\frac{f\left(-\xi_{\mathbf{K}}\right)}{E_{\mathbf{p};\mathbf{k}_{1}\mathbf{k}_{2}\cdots\mathbf{K};\mathbf{q}_{1}\mathbf{q}_{2}\cdots\mathbf{q}_{n}}^{(n)}}\right]^{-1}\left[\sum_{j=1}^{n}(-1)^{j-1}\alpha_{\mathbf{q}_{1}\mathbf{q}_{2}\cdots\mathbf{q}_{n-j}\mathbf{q}_{n-j+2}\cdots\mathbf{q}_{n}}^{\mathbf{k}_{1}\mathbf{k}_{2}\cdots\mathbf{k}_{n-1}}+\right.\nonumber \\
 &  & \left.\sum_{\mathbf{K}}\frac{\sum_{i=1}^{n-1}G_{\mathbf{q}_{1}\mathbf{q}_{2}\cdots\mathbf{q}_{n}}^{\mathbf{k}_{1}\mathbf{k}_{2}\cdots\mathbf{k}_{n-i-1}\mathbf{K}\mathbf{k}_{n-i+1}\cdots\mathbf{k}_{n-1}}}{E_{\mathbf{p};\mathbf{k}_{1}\mathbf{k}_{2}\cdots\mathbf{K};\mathbf{q}_{1}\mathbf{q}_{2}\cdots\mathbf{q}_{n}}^{(n)}}f\left(-\xi_{\mathbf{K}}\right)-\sum_{\mathbf{K}\mathbf{Q}}\frac{G_{\mathbf{q}_{1}\mathbf{q}_{2}\cdots\mathbf{q}_{n}\mathbf{Q}}^{\mathbf{k}_{1}\mathbf{k}_{2}\cdots\mathbf{k}_{n-1}\mathbf{K}}}{E_{\mathbf{p};\mathbf{k}_{1}\mathbf{k}_{2}\cdots\mathbf{K};\mathbf{q}_{1}\mathbf{q}_{2}\cdots\mathbf{q}_{n}}^{(n)}}f\left(-\xi_{\mathbf{K}}\right)f\left(\xi_{\mathbf{Q}}\right)\right].\label{eq:CAnph_U0}
\end{eqnarray}
In the first term on the right-hand-side of the equation, only the
sub-terms with a fixed set of indices $\mathbf{k}_{1}\mathbf{k}_{2}\cdots\mathbf{k}_{n-1}$
survive. When we exchange any two indices in the set $\{\mathbf{q}\}=\mathbf{q}_{1}\mathbf{q}_{2}\cdots\mathbf{q}_{n}$,
the sign $(-1)^{j-1}$ ensures that the first term is antisymmetrized,
see, for example, Appendix A. In each sub-term of the second term
$\sum_{\mathbf{K}}(\cdots)$, we will find a sign factor $(-1)^{i-1}$,
if we move the momentum $\mathbf{K}$ to the right-hand-side of $\mathbf{k}_{n-1}$.
This sign factor $(-1)^{i-1}$ has the similar effect for antisymmetrization.
When we exchange any two indices in the set $\{\mathbf{k}\}=\mathbf{k}_{1}\mathbf{k}_{2}\cdots\mathbf{k}_{n-1}$,
the second term that consists of $(n-1)$ sub-terms is antisymmetrized.
Moreover, there is only one item in the last term, which involves
a higher order $G_{\mathbf{q}_{1}\mathbf{q}_{2}\cdots\mathbf{q}_{n}\mathbf{Q}}^{\mathbf{k}_{1}\mathbf{k}_{2}\cdots\mathbf{k}_{n-1}\mathbf{K}}$.
We will make the coupled equations enclosed, if we drop this last
term and truncate to the level of $n$-th particle-hole excitations.

As a concrete example, up to the level of three-particle-hole excitations,
we obtain the coupled equations (by default, we take $\mathbf{k}=\mathbf{k}_{1},\mathbf{k}'=\mathbf{k}_{2},\mathbf{k}''=\mathbf{k}_{3}$
and $\mathbf{q}=\mathbf{q}_{1},\mathbf{q}'=\mathbf{q}_{2},\mathbf{q}''=\mathbf{q}_{3}$),
\begin{eqnarray}
F\left(\mathbf{q}\right) & = & \left[\frac{1}{U}+\sum_{\mathbf{k}}\frac{f\left(-\xi_{\mathbf{k}}\right)}{E_{\mathbf{p};\mathbf{k};\mathbf{q}}^{(1)}}\right]^{-1}\left[1-\sum_{\mathbf{kq'}}\frac{G_{\mathbf{qq}'}^{\mathbf{k}}}{E_{\mathbf{p};\mathbf{k};\mathbf{q}}^{(1)}}f\left(-\xi_{\mathbf{k}}\right)f\left(\xi_{\mathbf{q'}}\right)\right],\label{eq:CA3phU0_E1}\\
G_{\mathbf{qq}'}^{\mathbf{k}} & = & \left[\frac{1}{U}+\sum_{\mathbf{k'}}\frac{f\left(-\xi_{\mathbf{k'}}\right)}{E_{\mathbf{p};\mathbf{kk'};\mathbf{qq'}}^{(2)}}\right]^{-1}\left[\left(\alpha_{\mathbf{q}}^{\mathbf{k}}-\alpha_{\mathbf{q'}}^{\mathbf{k}}\right)+\sum_{\mathbf{k'}}\frac{G_{\mathbf{qq}'}^{\mathbf{k'}}}{E_{\mathbf{p};\mathbf{kk'};\mathbf{qq'}}^{(2)}}f\left(-\xi_{\mathbf{k'}}\right)-\sum_{\mathbf{k'q''}}\frac{G_{\mathbf{qq}'\mathbf{q}''}^{\mathbf{kk}'}}{E_{\mathbf{p};\mathbf{kk'};\mathbf{qq'}}^{(2)}}f\left(-\xi_{\mathbf{k'}}\right)f\left(\xi_{\mathbf{q''}}\right)\right],\label{eq:CA3phU0_E2}\\
G_{\mathbf{qq}'\mathbf{q}''}^{\mathbf{kk}'} & = & \left[\frac{1}{U}+\sum_{\mathbf{k''}}\frac{f\left(-\xi_{\mathbf{k''}}\right)}{E_{\mathbf{p};\mathbf{kk'k''};\mathbf{qq'q''}}^{(3)}}\right]^{-1}\left[\left(\alpha_{\mathbf{q}\mathbf{q}'}^{\mathbf{kk'}}-\alpha_{\mathbf{qq''}}^{\mathbf{kk'}}+\alpha_{\mathbf{q'q''}}^{\mathbf{kk'}}\right)+\sum_{\mathbf{k''}}\frac{\left(G_{\mathbf{qq}'\mathbf{q}''}^{\mathbf{kk}''}-G_{\mathbf{qq}'\mathbf{q}''}^{\mathbf{k'k}''}\right)}{E_{\mathbf{p};\mathbf{kk'k''};\mathbf{qq'q''}}^{(3)}}f\left(-\xi_{\mathbf{k''}}\right)\right].\label{eq:CA3phU0_E3}
\end{eqnarray}
\end{widetext}Here, in the first equation Eq. (\ref{eq:CA3phU0_E1})
we would like to keep the notation of $F(\mathbf{q})$, instead of
using $G_{\mathbf{q}}$. In the last equation Eq. (\ref{eq:CA3phU0_E3}),
we have dropped the last term that is related to the higher order
$G_{\mathbf{qq}'\mathbf{q}''\mathbf{q}'''}^{\mathbf{kk}'\mathbf{k}''}$
(or more precisely, $G_{\mathbf{qq}'\mathbf{q}''\mathbf{Q}}^{\mathbf{kk}'\mathbf{K}}$).

\section{Diagrammatic theory versus Chevy ansatz}

It is readily seen that Eq. (\ref{eq:GammaNP1}) from the diagrammatic
theory and Eq. (\ref{eq:CAnph_U0}) from the Chevy ansatz approach
have the same structure. Moreover, when we take the on-shell values
for $\{k_{l}\}$ and $\{q_{l}\}$, by using the explicit expression
of the many-body $T$-matrix $T_{2}$, Eq. (\ref{eq: T2}), we immediately
identify ($\omega=\tilde{E}=E-E_{\textrm{FS}}-\nu U$),
\begin{equation}
T_{2}\left(p+\sum_{l}q_{l}-\sum_{l\neq n}k_{l}\right)=\left[\frac{1}{U}+\sum_{\mathbf{K}}\frac{f\left(-\xi_{\mathbf{K}}\right)}{E_{\mathbf{p};\mathbf{k}_{1}\cdots\mathbf{K};\mathbf{q}_{1}\cdots\mathbf{q}_{n}}^{(n)}}\right]^{-1}.
\end{equation}
As both Eq. (\ref{eq:GammaNP1}) and Eq. (\ref{eq:CAnph_U0}) are
derived in an exact manner under the same conditions, we conclude
that they must be the same equation. By comparing the corresponding
terms in the two equations, therefore, we should have the relations,
\begin{eqnarray}
G_{\mathbf{q}_{1}\mathbf{q}_{2}\cdots\mathbf{q}_{n}}^{\mathbf{k}_{1}\mathbf{k}_{2}\cdots\mathbf{k}_{n-1}} & = & \Gamma_{n+1}\left(\{k_{l}\}_{l\neq n};p,\{q_{l}\}\right),\label{eq: RelationG}\\
\alpha_{\mathbf{q}_{1}\mathbf{q}_{2}\cdots\mathbf{q}_{n-j}\mathbf{q}_{n-j+2}\cdots\mathbf{q}_{n}}^{\mathbf{k}_{1}\mathbf{k}_{2}\cdots\mathbf{k}_{n-1}} & = & \left(-1\right)^{j-1}A_{j},\label{eq: RelationAlpha0}
\end{eqnarray}
where in the multi-particle vertex function $\Gamma_{n+1}$ and $A_{j}$,
we need to take the on-shell values for all the four-momenta $\{k_{l}\}$
and $\{q_{l}\}$. Now, we would like to claim that the second relation
Eq. (\ref{eq: RelationAlpha0}) is more fundamental than the first
relation Eq. (\ref{eq: RelationG}), in the sense that we can derive
Eq. (\ref{eq: RelationG}) by using Eq. (\ref{eq: RelationAlpha0}).
To see this, let us rewrite Eq. (\ref{eq: RelationAlpha0}) into the
form,
\begin{equation}
\alpha_{\mathbf{q}_{1}\mathbf{q}_{2}\cdots\mathbf{q}_{n}}^{\mathbf{k}_{1}\mathbf{k}_{2}\cdots\mathbf{k}_{n}}=-\frac{\Gamma_{n+1}\left(\{k_{l}\}_{l\neq n};p,\{q_{l}\}\right)}{E_{\mathbf{p};\{\mathbf{k}\};\{\mathbf{q}\}}^{(n)}},\label{eq:RelationAlpha}
\end{equation}
where we have used the fact that,
\begin{equation}
G_{0\downarrow}\left(p+\sum_{l}q_{l}-\sum_{l}k_{l}\right)=-\frac{1}{E_{\mathbf{p};\{\mathbf{k}\};\{\mathbf{q}\}}^{(n)}}.
\end{equation}
As we emphasized earlier, $\Gamma_{n+1}(\{k_{l}\}_{l\neq n};p,\{q_{l}\})$
has a unique feature that it is independent on the four-momentum $k_{n}=(\mathbf{k}_{n},\xi_{\mathbf{k}_{n}})$.
The $\mathbf{k}_{n}$-dependence of $\alpha_{\mathbf{q}_{1}\mathbf{q}_{2}\cdots\mathbf{q}_{n}}^{\mathbf{k}_{1}\mathbf{k}_{2}\cdots\mathbf{k}_{n}}$
therefore only comes from $G_{0\downarrow}(p+\sum_{l}q_{l}-\sum_{l}k_{l})$
or $E_{\mathbf{p};\{\mathbf{k}\};\{\mathbf{q}\}}^{(n)}$. This feature
has an interesting consequence, if we calculate $G_{\mathbf{q}_{1}\mathbf{q}_{2}\cdots\mathbf{q}_{n}}^{\mathbf{k}_{1}\mathbf{k}_{2}\cdots\mathbf{k}_{n-1}}$
by using its definition,\begin{widetext}
\begin{equation}
G_{\mathbf{q}_{1}\mathbf{q}_{2}\cdots\mathbf{q}_{n}}^{\mathbf{k}_{1}\mathbf{k}_{2}\cdots\mathbf{k}_{n-1}}\equiv U\sum_{\mathbf{k}_{n}}\alpha_{\mathbf{q}_{1}\mathbf{q}_{2}\cdots\mathbf{q}_{n}}^{\mathbf{k}_{1}\cdots\mathbf{k}_{n-1}\mathbf{k}_{n}}f\left(-\xi_{\mathbf{k}_{n}}\right)=\left[U\sum_{\mathbf{k}_{n}}\frac{\Gamma_{n+1}\left(\{k_{l}\}_{l\neq n};p,\{q_{l}\}\right)f\left(-\xi_{\mathbf{k}_{n}}\right)}{\left(\omega+\sum_{l}\xi_{\mathbf{q}_{l}}-\sum_{l\neq n}\xi_{\mathbf{k}_{l}}\right)-\xi_{\mathbf{k}_{n}}-\varepsilon_{\mathbf{p}-\mathbf{P}_{\vec{\kappa}_{n}}}^{I}}\right].
\end{equation}
\end{widetext}The integral over $\mathbf{k}_{n}$ in the above equation
diverges and when it multiplies with the vanishingly small $U$, we
obtain $1$, due to the regularization relation Eq. (\ref{eq: Ureg}).
Thus, we immediately recover Eq. (\ref{eq: RelationG}), as promised.

To directly show the usefulness of the two relations, Eq. (\ref{eq:RelationAlpha})
and Eq. (\ref{eq: RelationG}), let us apply them to the four-particle
vertex function $\Gamma_{4}(kk';p,qq'q'')$ with $n=3$. From Fig.
\ref{fig6: Gamma4}, we may write down the following expression for
the diagrams:
\begin{equation}
\frac{\Gamma_{4}\left(kk';p,qq'q''\right)}{T_{2}\left(p+q+q'+q''-k-k'\right)}=\sum_{j=1}^{3}A_{j}+\sum_{i=1}^{2}B_{i}+C,\label{eq:Diagram3ph}
\end{equation}
where the expressions of $A_{j}$ ($j=1,2,3$) are,
\begin{eqnarray}
A_{1} & = & +G_{0\downarrow}\left(p+q+q'-k-k'\right)\Gamma_{3}\left(k;p,qq'\right),\\
A_{2} & = & -G_{0\downarrow}\left(p+q+q''-k-k'\right)\Gamma_{3}\left(k;p,qq''\right),\\
A_{3} & = & +G_{0\downarrow}\left(p+q'+q''-k-k'\right)\Gamma_{3}\left(k;p,q'q''\right),
\end{eqnarray}
and the expressions of $B_{1}$, $B_{2}$ and $C$, after the Matsubara
frequency summation, are given by,\begin{widetext}
\begin{eqnarray}
B_{1} & = & -\sum_{\mathbf{k}''}G_{0\downarrow}\left(p+q+q'+q''-k-k'-k''\right)\Gamma_{4}\left(kk'';p,qq'q''\right)f\left(-\xi_{\mathbf{k}''}\right),\\
B_{2} & = & -\sum_{\mathbf{k}''}G_{0\downarrow}\left(p+q+q'+q''-k-k'-k''\right)\Gamma_{4}\left(k''k';p,qq'q''\right)f\left(-\xi_{\mathbf{k}''}\right),\\
C & = & +\sum_{\mathbf{k}''\mathbf{q}'''}G_{0\downarrow}\left(p+q+q'+q''-k-k'-k''\right)\Gamma_{5}\left(kk'k'';p,qq'q''q'''\right)f\left(-\xi_{\mathbf{k}''}\right)f\left(\xi_{\mathbf{q}'''}\right).
\end{eqnarray}
It is straightforward to apply Eq. (\ref{eq:RelationAlpha}) to $A_{j}$
and to verify $A_{1}=+\alpha_{\mathbf{q}\mathbf{q}'}^{\mathbf{kk'}}$,
$A_{2}=-\alpha_{\mathbf{qq''}}^{\mathbf{kk'}}$, and $A_{3}=+\alpha_{\mathbf{q'q''}}^{\mathbf{kk'}}$,
which are the first three terms in the square bracket of the right-hand-side
of Eq. (\ref{eq:CA3phU0_E3}). Moreover, by using the following relation
with on-shell momenta,
\begin{equation}
G_{0\downarrow}\left(p+q+q'+q''-k-k'-k''\right)=-\frac{1}{E_{\mathbf{p};\mathbf{kk'k''};\mathbf{qq'q''}}^{(3)}},
\end{equation}
\end{widetext}and by applying Eq. (\ref{eq: RelationG}) in $B_{i}$
to replace $\Gamma_{4}$, we find that the last two terms in the square
bracket of Eq. (\ref{eq:CA3phU0_E3}) correspond to $B_{1}$ and $B_{2}$,
respectively. It is also not difficult to verify that, the dropped
higher order term in Eq. (\ref{eq:CA3phU0_E3}) is given by $C$.
In this way, we directly reproduce the Chevy ansatz result of Eq.
(\ref{eq:CA3phU0_E3}), by applying the diagrammatic theory. 

\section{Fermi polarons in one-dimensional lattices}

We now turn to consider the numerical calculations of the polaron
spectral function, beyond the commonly-used approximation of including
just one-particle-hole excitations. However, at finite temperature
there is a serious numerical problem related to the zeros of the excitation
energy $E_{\mathbf{p};\{\mathbf{k}\};\{\mathbf{q}\}}^{(n)}$, which
create a lot of singularities (i.e., poles) in the coefficients $\alpha_{\mathbf{q}_{1}\mathbf{q}_{2}\cdots\mathbf{q}_{n}}^{\mathbf{k}_{1}\mathbf{k}_{2}\cdots\mathbf{k}_{n}}$
and $G_{\mathbf{q}_{1}\mathbf{q}_{2}\cdots\mathbf{q}_{n}}^{\mathbf{k}_{1}\mathbf{k}_{2}\cdots\mathbf{k}_{n-1}}$,
as can be clearly seen from the relation Eq. (\ref{eq:RelationAlpha}).
These singularities make it impossible to directly calculate the various
integrals appearing in the exact set of the coupled equations, Eq.
(\ref{eq:GammaNP1}) or Eq. (\ref{eq:CAnph_U0}). This numerical problem
exists even at zero temperature, if we want to study the repulsive
polaron branch at positive energy.

To overcome the numerical difficulty, we may introduce a finite broadening
factor $\eta$ to the frequency (i.e., $\omega\rightarrow\omega_{\eta}\equiv\omega+i\eta$).
Here, for simplicity, we focus on the case of one-dimensional lattices,
where the value of momentum is restricted to the first Brillouin zone.
We will consider the inclusion of two-particle-hole excitations. A
slight inconvenience is that the on-site interaction strength $U$
is nonzero. Therefore, we must solve the coupled equations Eq. (\ref{eq:CA2ph_Fq}),
Eq. (\ref{eq:CA2ph_Rk}), Eq. (\ref{eq:CA2ph_Gkqqp}) and Eq. (\ref{eq:CA2ph_Hkkpq}),
with momentum $k\subseteq[-\pi,+\pi]$ and $q\subseteq[-\pi,+\pi]$
at a finite broadening factor $\eta$. Eventually, we extrapolate
$\eta$ to zero and obtain the $\eta$-independent polaron self-energy
in Eq. (\ref{eq:CA2ph_SelfEnergy}) and hence the polaron spectral
function. 

In future studies, this numerical trick might be extended to the three-dimensional
free space, with some improvements. With more elaborate numerical
efforts, then we may calculate the more interesting finite-temperature
spectral function of a unitary Fermi polaron.

\subsection{Numerical procedure}

In one-dimensional lattices, we take $\varepsilon_{k}=-2t\cos k+2t$
and $\varepsilon_{p}^{I}=-2t_{d}\cos p+2t_{d}$. Typically, we set
the energy scale $t=1$. The hopping strength of the impurity $t_{d}$
is determined by the mass of the impurity, since $t_{d}/t=m/m_{I}$.
At finite temperature, the chemical potential $\mu$ is to be fixed
by the filling factor $\nu$. We perform numerical calculations for
a given momentum $p$ and a given energy $\omega_{\eta}=\omega+i\eta$,
so $E_{p;k;q}^{(1)}=-\omega_{\eta}+\varepsilon_{p+q-k}^{I}+\varepsilon_{k}-\varepsilon_{q}$
and $E_{p;kk';qq'}^{(2)}=-\omega_{\eta}+\varepsilon_{p+q+q'-k-k'}^{I}+\varepsilon_{k}+\varepsilon_{k'}-\varepsilon_{q}-\varepsilon_{q'}$.
Our numerical procedure consists of the following three steps.
\begin{itemize}
\item \uline{Step 1}. For a given $J(k,q)$, which initially is zero,
we iteratively solve the coupled equations Eq. (\ref{eq:CA2ph_Fq})
and Eq. (\ref{eq:CA2ph_Rk}). To check convergence, we compare $F(q)$
with a previously saved $F(q)$ (which is zero from the beginning
as well). If the difference is smaller than a certain criterion (i.e.,
the average difference in relative is smaller than $10^{-8}$), we
jump to Step 3; otherwise, continue with Step 2.
\item \uline{Step 2}. For $F(q)$ and $R(k)$ generated from Step 1,
we iteratively solve the coupled equations Eq. (\ref{eq:CA2ph_Gkqqp})
and Eq. (\ref{eq:CA2ph_Hkkpq}) for $G_{qq'}^{k}$ and $H_{q}^{kk'}$.
As this is a very time-consuming procedure, we do not require the
full convergence and the iteration lasts for a few times (we typically
choose 8 times). Also, since $G_{q'q}^{k}=-G_{qq'}^{k}$ and $H_{q}^{k'k}=-H_{q}^{kk'}$,
we only need to calculate the case with $q'<q$ and $k'<k$. In each
interaction, $J(k,q)$ will be updated. We use the two ways to calculate
$J(k,q)$, by using $G_{qq'}^{k}$ and $H_{q}^{kk'}$, and monitor
their difference. If the difference continues to decrease, the iteration
moves on the correct direction towards convergence. We then go back
to Step 1, with the updated $J(k,q)$. It is important to note that,
in each iteration, we do not entirely replace $G_{qq'}^{k}$ and $H_{q}^{kk'}$
with the new values (resulting from Eq. (\ref{eq:CA2ph_Gkqqp}) and
Eq. (\ref{eq:CA2ph_Hkkpq})). Instead, we only mix a small portion
of the new values for the update, for example, $10\%$ of the new
values. This treatment effectively removes the possible nonlinearity
occurring in the iteration procedure.
\item \uline{Step 3}. At this step, we obtain the converged $F(q)$ or
$R(k)$, we then calculate the self-energy by using Eq. (\ref{eq:CA2ph_SelfEnergy})
and the spectral function of Fermi polarons.
\end{itemize}
For the numerical integration, we typically use the $96$-point gaussian
quadrature approach, by discretizing the momentum $k$ or $q$ in
the range $[-\pi,+\pi]$. For this grid size, it costs a few minutes
to finish the iterations in Step 2. For a single set of given values
$p$ and $\omega_{\eta}$, the whole numerical iteration procedure
costs about an hour. We will perform the calculations for three values
of the broadening factor $\eta\sim t$. A good choice seems to be
$\eta=0.6t$, $0.9t$ and $1.2t$. We then extrapolate the values
of the self-energy to the limit $\eta=0^{+}$, by fitting them to
a quadratic function.

At zero temperature, we must take care of the sharp Fermi surface.
In this case, we know that the hole state is given by $-k_{F}<q<k_{F}$
and the particle state satisfies $k_{F}<k<2\pi-k_{F}$ (or equivalently
$k$ is the range $(-\pi,-k_{F})\cup(k_{F},+\pi)$), where the Fermi
momentum $k_{F}=\nu\pi$. Therefore, we can discretize the grid points
in $(-k_{F},k_{F})$ for $q$ and $(k_{F},2\pi-k_{F})$ for $k$.

\begin{figure}
\begin{centering}
\includegraphics[width=0.5\textwidth]{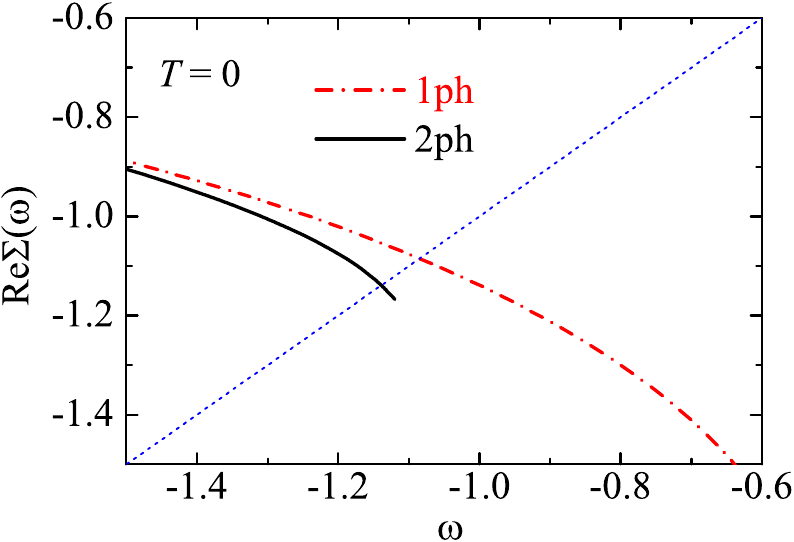}
\par\end{centering}
\caption{\label{fig9: selfenergyT0} The self-energy $\Sigma(\mathbf{p}=0,\omega)$
of Fermi polarons in 1D lattices at zero momentum, without the broadening
factor (i.e., $\eta=0$). The frequency $\omega$ is in the range
for the attractive polaron branch. The black solid lines and red dash-dotted
lines show the results with one-particle-hole (1ph) excitations and
with two-particle-hole (2ph) excitations, respectively. The blue dotted
straight line is $\omega-\varepsilon_{\mathbf{p}=0}^{I}$. It crosses
with the self-energy at the energy of the attractive polaron $\mathcal{E}_{A}$.
We find $\mathcal{E}_{A}\simeq-1.14t$ if we take into account two-particle-hole
excitations. Both the self-energy $\Sigma$ and the frequency $\omega$
are measured in units of the hopping amplitude of spin-up fermions
$t$. Throughout the work, we consider a lattice filling factor $\nu=0.2$,
the on-site interaction strength $U=-4t$, and an impurity hopping
amplitude $t_{d}=t$.}
\end{figure}

\begin{figure}
\begin{centering}
\includegraphics[width=0.5\textwidth]{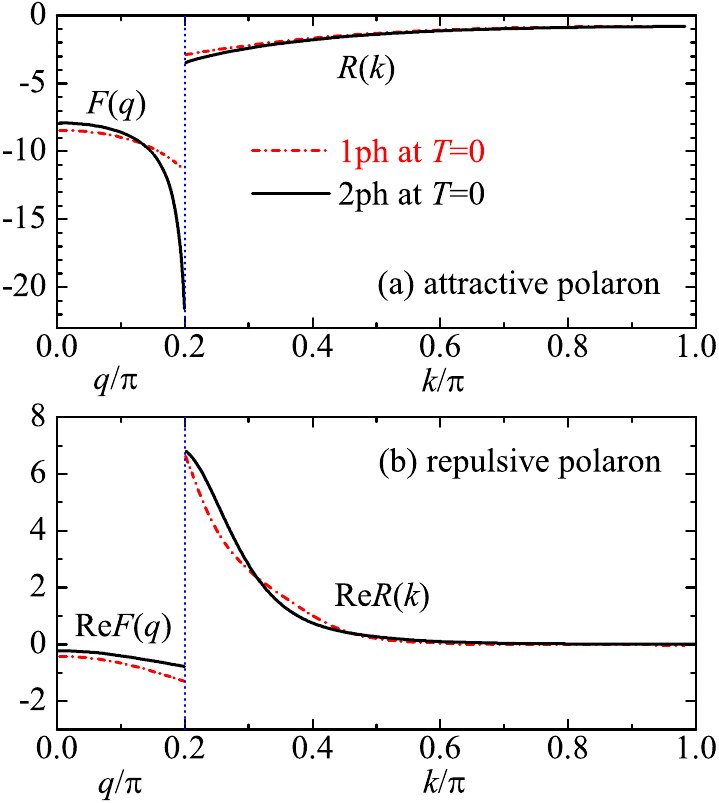}
\par\end{centering}
\caption{\label{fig10: FqRkT0} The two functions $F(q)$ and $R(k)$ at zero
temperature, where we can restrict $q\protect\leq\nu\pi$ and $k\protect\geq(1-\nu)\pi$.
Two frequencies are considered: $\omega=\mathcal{E}_{A}\simeq-1.14t$
at the attractive polaron energy (a) and $\omega=\mathcal{E}_{R}\simeq1.54t$
at the repulsive polaron energy (b). The black solid lines and red
dash-dotted lines show the results with one-particle-hole (1ph) excitations
and with two-particle-hole (2ph) excitations, respectively. In (a)
for the attractive polaron branch, we set $\eta=0$. In contrast,
in (b) for the repulsive polaron branch, we introduce a broadening
factor $\eta=0.6t$, to avoid the numerical singularity and instability.}
\end{figure}

\subsection{Results and discussions}

In our numerical calculations, we consider an equal mass of the spin-up
fermions and the impurity, so $t_{d}=t=1$. We fix the filling factor
$\nu=0.2$ and take an interaction strength $U=-4t$. We also set
the polaron momentum to be zero, $p=0$. At these parameters, we find
that the energies of the attractive polaron and the repulsive polaron
are roughly given by, $\mathcal{E}_{A}\sim-t$ and $\mathcal{E}_{R}\sim+t$,
respectively. We are particularly interested in the effects of two-particle-hole
excitations on $F(q)$ and $R(k)$, and consequently, the resulting
improvement to the polaron self-energy and spectral function.

\subsubsection{$F(q)$ and $R(k)$ at zero temperature}

Let us first consider the zero-temperature case. At $T=0$, one advantage
is that the attractive polaron is the unique ground-state of the quantum
many-body system, so the excitation energies $E_{p;k;q}^{(1)}>0$
and $E_{p;kk';qq'}^{(2)}>0$ at $\omega\sim\mathcal{E}_{A}\sim-t$.
Thus, there is no singularity in the coupled equations for the attractive
polaron branch. We do not need to introduce the small broadening factor
$\eta$. In Fig. \ref{fig9: selfenergyT0}, we show the polaron self-energy
$\Sigma(\omega$) at zero momentum $p=0$, calculated with the inclusion
of one-particle-hole (1ph) and two-particle-hole (2ph) excitations,
strictly at $\eta=0$. We find $\mathcal{E}_{A}\simeq-1.14t$ with
2ph excitations, which is slightly smaller than the prediction with
1ph excitations, as expected. We note that, at $\eta=0$ the numerical
calculations with 2ph excitations have to stop at a threshold energy
slightly larger than $\mathcal{E}_{A}$, above which we encounter
the zeros of $E_{p;kk';qq'}^{(2)}$ and therefore our numerical procedure
becomes unstable.

In Fig. \ref{fig10: FqRkT0}, we plot the curves $F(q)$ and $R(k)$,
predicted with 1ph and 2ph excitations, at the attractive polaron
energy $\omega=\mathcal{E}_{A}\simeq-1.14t$ (a) and at the repulsive
polaron energy $\omega=\mathcal{E}_{R}\simeq1.54t$ (b). For the latter,
we must include a spectral broadening factor (i.e., $\eta=0.6t$)
to ensure the numerical stability, and therefore $F(q)$ and $R(k)$
become complex. Both $F(q)$ and $R(k)$ are even functions, i.e.,
$F(-q)=-F(q)$ and $R(-k)=-R(k)$. 

At the attractive polaron energy in Fig. \ref{fig10: FqRkT0}(a),
we observe that $F(q)$ at $\eta=0$ calculated with 2ph excitations
decreases very rapidly when $q$ approaches the Fermi point, $q\rightarrow k_{F}$.
This singular behavior might be removed by either a finite broadening
factor $\eta$ or a nonzero temperature $T$. At the repulsive polaron
energy in Fig. \ref{fig10: FqRkT0}(b), we find that $R(k)$ has a
much larger magnitude than $F(q)$. This finding is probably not expected,
since $R(k)$ has to vanish identically in the free space with an
infinitesimal interaction strength $U\rightarrow0^{-}$.

\begin{figure}
\begin{centering}
\includegraphics[width=0.5\textwidth]{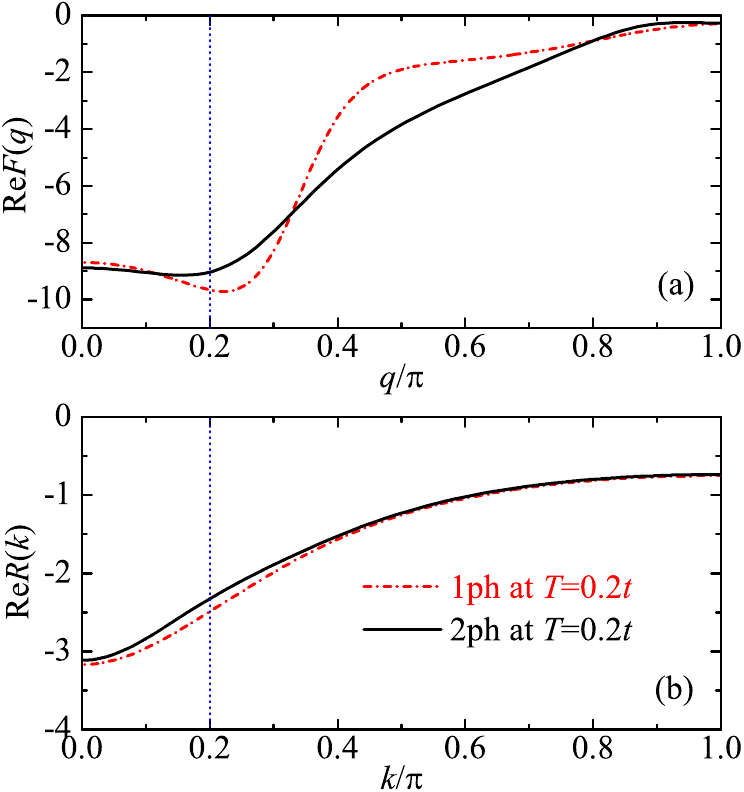}
\par\end{centering}
\caption{\label{fig11: FqRkT02t} The real part of $F(q)$ (a) and $R(k)$
(b) at $\omega=\mathcal{E}_{A}\simeq-1.14t$ and at nonzero temperature
$T=0.2t$, where the range of $k$ and $q$ extends to the whole Brillouin
zone. At this temperature, we take a broadening factor $\eta=0.6t$,
to avoid the numerical instability. The black solid lines and red
dash-dotted lines show the results with one-particle-hole (1ph) excitations
and with two-particle-hole (2ph) excitations, respectively.}
\end{figure}

\begin{figure}
\begin{centering}
\includegraphics[width=0.5\textwidth]{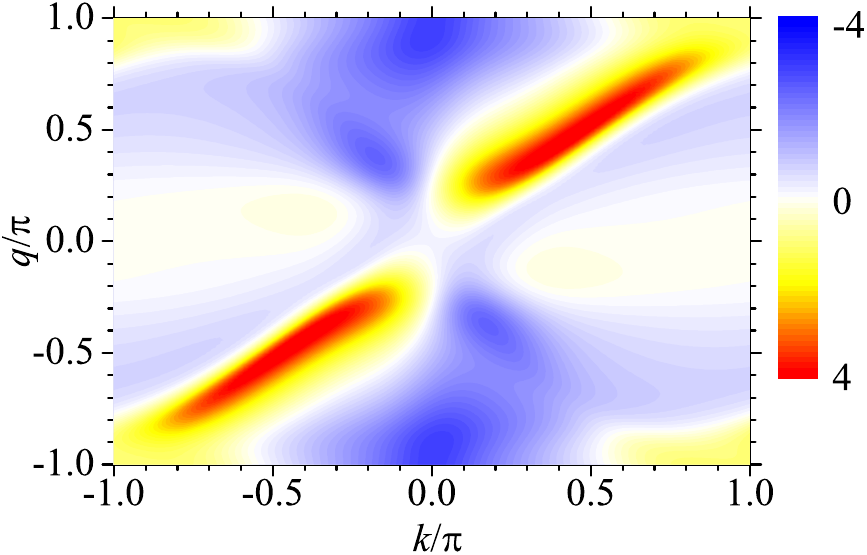}
\par\end{centering}
\caption{\label{fig12: JkqT02t} A contour plot of the function $\textrm{Re}J(k,q)$
at $\omega=\mathcal{E}_{A}\simeq-1.14t$ and at the temperature $T=0.2t$,
with its magnitude indicated by a color bar. It is easy to see that
$J(k,q)$ has an odd parity, i.e., $J(-k,-q)=-J(k,q)$. We take a
broadening factor $\eta=0.6t$, to avoid the numerical instability.}
\end{figure}

\subsubsection{$F(q)$ and $R(k)$ at nonzero temperature}

Let us now turn to investigate the finite-temperature polaron spectral
function. In this case, the range of $k$ and $q$ in the functions
$F(q)$ and $R(k)$ extends to the whole Brillouin zone $[-\pi,+\pi]$,
as given in Fig. \ref{fig11: FqRkT02t}, for the attractive Fermi
polaron at $T=0.2t$ with a broadening factor $\eta=0.6t$. For $F(q)$
in Fig. \ref{fig11: FqRkT02t}(a), we find a large difference in the
predictions with 1ph and 2ph excitations, at $q\sim0.5\pi$. However,
this difference may not lead to too much difference in the calculated
self-energy $\Sigma(\omega)$, due to the existence of the thermal
weighting factor of the Fermi distribution function $f(\xi_{q})$,
see, for example, Eq. (\ref{eq:CA2ph_SelfEnergy}).

To understand why we obtain a very different $F(q)$ with and without
2ph excitations, in Fig. \ref{fig12: JkqT02t} we show a contour plot
of the corresponding $\textrm{Re}J(k,q)$, which measures the importance
of the 2ph functions $G_{qq'}^{k}$ and $H_{q}^{kk'}$. In contrast
to $F(q)$ and $R(k)$, $J(k,q)$ has an odd parity. Therefore, at
the origin $k=0$ and $q=0$, $J(k,q)$ is strictly zero. We find
that $J(k,q)$ is a rapidly varying function as a function of either
$k$ and $q$. This observation agrees with our impression that $G_{qq'}^{k}$
and $H_{q}^{kk'}$ could be a very singular function due to the smallness
of the excitation energy $E_{p;kk';qq'}^{(2)}$. We also find that
$J(k,q)$ has a large magnitude roughly along the diagonal direction
$k=q$, peaking at about $k=q\sim\pm0.5\pi$. As $J(k,q)$ is an input
to Eq. (\ref{eq:CA2ph_Fq}) and Eq. (\ref{eq:CA2ph_Rk}), the existence
of these peaks may qualitatively explain why the $F(q)$ calculated
with and without 2ph excitations show a large difference.

\begin{figure}
\begin{centering}
\includegraphics[width=0.5\textwidth]{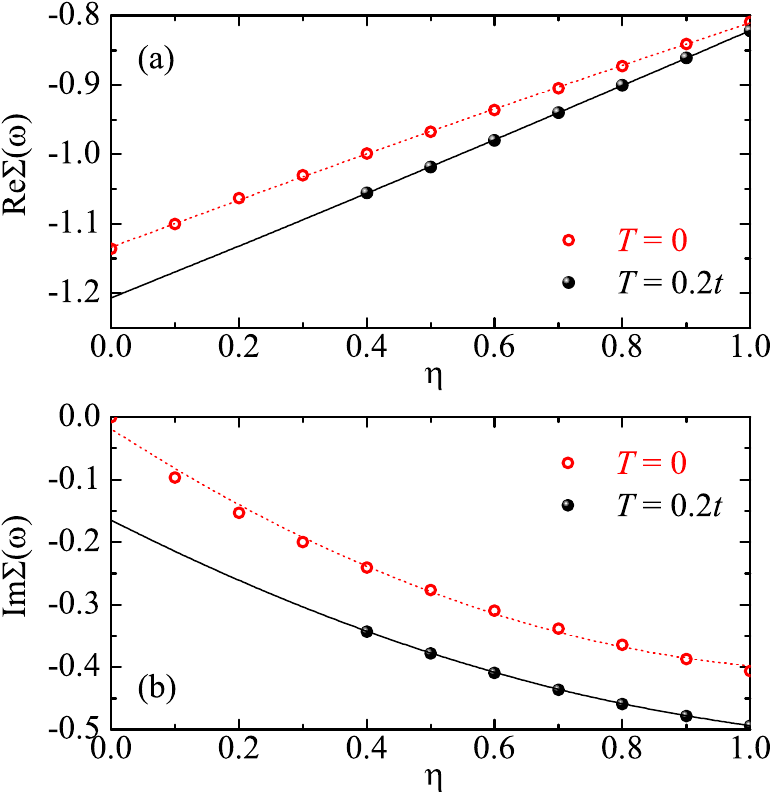}
\par\end{centering}
\caption{\label{fig13: YitaDependence} The real part (a) and imaginary part
(b) of the polaron self-energy at $\omega=\mathcal{E}_{A}\simeq-1.14t$,
as a function of the broadening factor $\eta$. The results with two-particle-hole
excitations are shown by red empty circles (for zero temperature)
and by black solid circles (for $T=0.2t$). The lines are the fits
to the results (symbols), with a quadratic function (i.e., a polynomial
of degree two). Both the self-energy $\Sigma$ and the broadening
factor $\eta$ are measured in units of the hopping amplitude of spin-up
fermions $t$.}
\end{figure}

\subsubsection{The $\eta$-dependence of the polaron self-energy}

Before we present the results on the $\eta$-independent polaron self-energy
and spectral function, it is useful to carefully check our extrapolating
strategy of a quadratic curve fitting. In Fig. \ref{fig13: YitaDependence},
we report the self-energy as a function of the broadening factor $\eta$
at zero temperature (empty red circles) and at $T=0.2t$ (black solid
circles), at the attractive polaron energy. The zero-temperature results
extend all the way down to $\eta=0^{+}$, due to the nonzero excitation
energy as we mentioned earlier. In contrast, at nonzero temperature,
our choice of the discretization grid for momentum (i.e., used for
the $96$-point gaussian quadrature integral) only allows us to accurately
calculate the self-energy with $\eta\geq0.4t$.

We observe that the results of the self-energy over a wide range of
$\eta$ can be well fitted by using a polynomial of degree two, at
both zero temperature and nonzero temperature. The fitting is particularly
accurate for the real part of the self-energy, with an absolute error
less than $0.01t$ (see Fig. \ref{fig13: YitaDependence}(a)). For
the imaginary part of the self-energy in Fig. \ref{fig13: YitaDependence}(b),
we roughly estimate that the absolute error of $\textrm{Im}\Sigma(\omega)$,
extrapolating to $\eta=0^{+}$, would be around a few percent of $t$.

\begin{figure}
\begin{centering}
\includegraphics[width=0.5\textwidth]{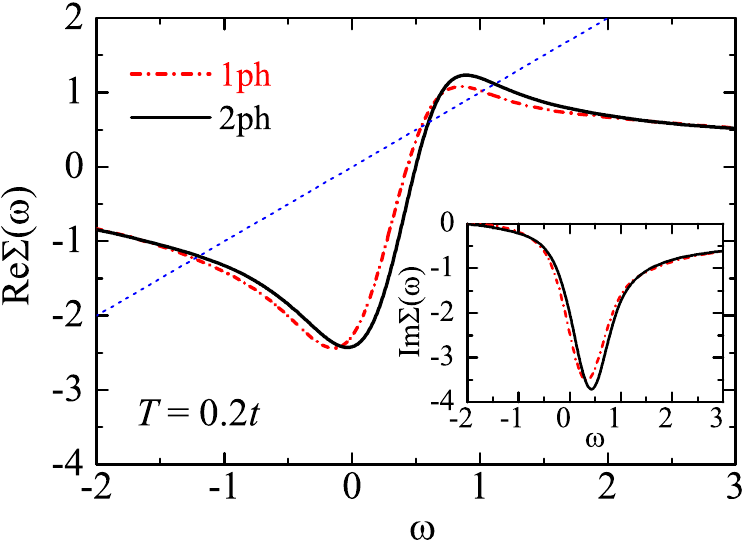}
\par\end{centering}
\caption{\label{fig14: selfenergyTx} The real part (main figure) and imaginary
part (inset) of the polaron self-energy at $T=0.2t$, calculated by
extrapolating $\eta$ to zero. The black solid lines and red dash-dotted
lines show the results with one-particle-hole (1ph) excitations and
with two-particle-hole (2ph) excitations, respectively.}
\end{figure}

\begin{figure}
\begin{centering}
\includegraphics[width=0.5\textwidth]{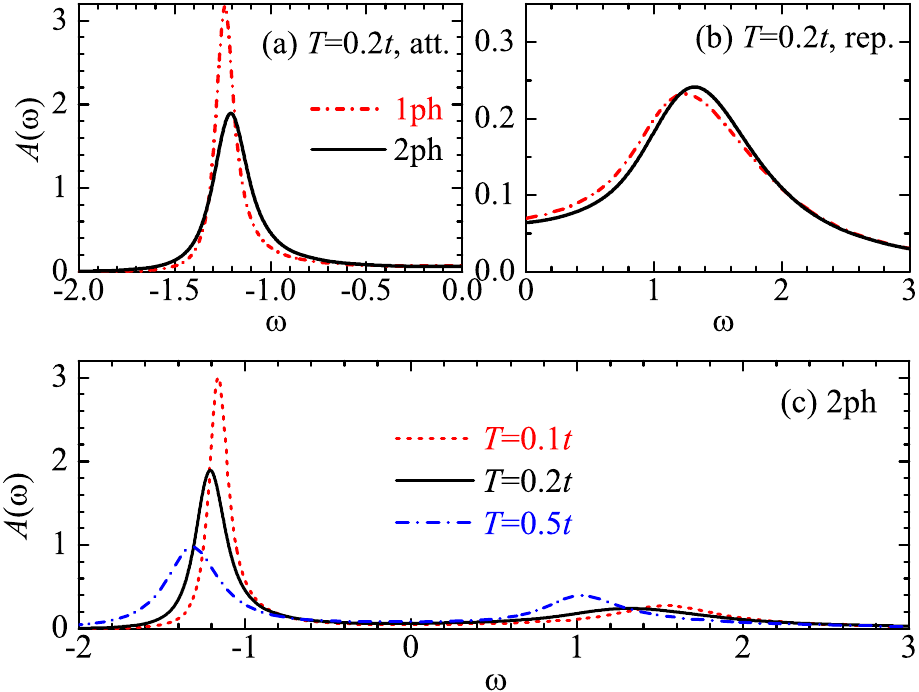}
\par\end{centering}
\caption{\label{fig15: akwTx} Upper panel: (a) and (b) show the polaron spectral
function at $T=0.2t$ near the attractive branch and the repulsive
branch, respectively. The black solid lines and red dash-dotted lines
show the results with one-particle-hole (1ph) excitations and with
two-particle-hole (2ph) excitations, respectively. Lower panel: the
polaron spectral function at $T=0.1t$ (red dashed line), $T=0.2t$
(black solid line), and $T=0.5t$ (blue dot-dashed line), calculated
by taking into account two-particle-hole excitations. All the results
are obtained by extrapolating $\eta$ to zero.}
\end{figure}

\subsubsection{The polaron self-energy and spectral function}

In Fig. \ref{fig14: selfenergyTx}, we show the polaron self-energy
at $T=0.2t$, obtained by extrapolating $\eta$ to zero, with the
inclusion of 1ph excitations (red dot-dashed lines) and 2ph excitations
(black solid lines). The corresponding polaron spectral function is
shown in Fig. \ref{fig15: akwTx}(a) and Fig. \ref{fig15: akwTx}(b),
for the attractive polaron branch and the repulsive polaron branch,
respectively. It is readily seen that, at nonzero temperature the
attractive polaron energy predicted with 2ph excitations (at $\mathcal{E}_{A}\simeq-1.20t$)
is slightly larger than the energy obtained with 1ph excitations (at
$\mathcal{E}_{A}\simeq-1.24t$). Therefore, the inclusion of more
particle-hole excitations does not necessarily make the attractive
polaron energy smaller, as we may naïvely anticipate from the viewpoint
that Chevy ansatz is a variational theory. As we mentioned earlier,
this is because, at nonzero temperature the attractive Fermi polaron
should be viewed as a collection of some many-body states, with different
weights. In our calculations with 2ph excitations, although the energies
of these (individual) many-body states become smaller due to the enlarged
Hilbert space, their weights are re-distributed. Roughly, the energy
of the many-body state with the maximum weight may then increase,
leading to the increase in the ``averaged'' attractive polaron energy.
On the other hand, we also find that the inclusion of 2ph excitations
enlarges the decay rate or the width of the attractive polaron. As
the consequence, the peak height of the attractive polaron decreases
significantly in the presence of 2ph excitations.

In Fig. \ref{fig15: akwTx}(c), we focus on the 2ph calculations and
investigate the temperature dependence of the polaron spectral function.
As the temperature increase, the energies of both attractive polaron
and repulsive polaron become smaller. For the attractive polaron,
the decay rate increases with temperature, so the peak height decreases.
In contrast, the peak height of the repulsive polaron shows a non-monotonic
dependence as a function of the temperature. It initially decreases
slightly at low temperature and then increases with increasing temperature.

\section{Conclusions and outlooks}

In conclusions, we have developed an exact theory of the spectral
function of Fermi polarons at finite temperature, by finding the complete
set of Feynman diagrams for the multi-particle vertex functions that
describe the multi-particle-hole excitations of the shake-up Fermi
sea. This is a rare case of quantum many-body theories, where the
exact solution is obtained by exhausting all the possible Feynman
diagrams for various vertex functions.

To understand why such an exact solution is feasible, we have provided
an alternative derivation, by generalizing the celebrated Chevy ansatz
to the finite temperature, with the inclusion of arbitrary numbers
of particle-hole excitations. We show that the ability to find an
exact solution roots in the closure of the Hilbert space for available
quantum states, in the case of a single impurity. 

We have rigorously prove that, for an infinitesimal interaction strength
in two-dimensional or three-dimensional free space, the diagrammatic
theory is fully equivalent to the Chevy ansatz approach, to any orders
of particle-hole excitations. In particular, the on-shell multi-particle
vertex functions are precisely the variational coefficients in the
Chevy ansatz. This remarkable relationship may provide a useful way
to calculate the multi-particle vertex functions, which are known
to be notoriously difficult to obtain in quantum many-body theories.

We have also shown that, to calculate the finite-temperature spectral
function of Fermi polarons for a nonzero interaction strength, the
Chevy ansatz is more powerful than the diagrammatic theory. In the
latter, the nonzero interaction strength leads to infinitely many
Feynman diagrams, whose roles are to be investigated in future studies.

To demonstrate the effects of multi-particle-hole excitations on the
finite-temperature spectral function, we have considered a specific
example of Fermi polarons in one-dimensional lattices, where the instability
of numerical calculations can be effectively removed. We have shown
that, for the attractive Fermi polaron at nonzero temperature, the
inclusion of two-particle-hole excitations typically leads to a larger
attractive polaron energy and a larger polaron decay rate. We have
explained that the larger polaron energy at finite temperature does
not contradict with the fact that the Chevy ansatz is a variational
approach for an individual quantum many-body state. However, the variational
viewpoint of the Chevy ansatz may not be worth emphasizing at nonzero
temperature, where the polaron state should be treated as a collection
of a number of quantum many-body states.

Our work can be straightforwardly generalized to handle the molecule
state of Fermi polarons, which is the ground state when the inter-particle
attraction becomes strong enough \citep{Punk2009,Mora2009,Combescot2009}.
In Appendix C, we list the set of equations for molecules, obtained
by using the Chevy ansatz approach. The parallel diagrammatic theory
will be described elsewhere.

In future studies, our work might be easily generalized to address
some long-standing problems in the polaron physics. The most interesting
example could be the polaron-polaron interaction \citep{Baroni2024}.
As we have emphasized, the closure of the Hilbert space for available
quantum states plays an important role to obtain the exact solution.
This closure of the Hilbert space should also hold for few impurities.
We should then be able to write down the generalized Chevy ansatz
for few impurities, in particular, for just two impurities. We may
also construct Feynman diagrams for the related multi-particle vertex
functions. In this way, we might be able to characterize the effective
polaron-polaron interaction, which is important to understand the
instability of a Fermi liquid of Fermi polarons at large impurity
concentration.

In addition, our approach could also be straightforwardly generalized
to investigate the spectral function of Bose polarons \citep{Scazza2022,Hu2016}
or crossover polarons \citep{Hu2022a} at finite temperature, where
the many-body environment is taken to be a weakly-interacting Bose
gas or a strongly interacting Fermi superfluid, respectively. In these
cases, special attention should be paid to the emergent three-body
bound states and the related three-body parameters. The possible construction
of relevant Feynman diagrams may provide insight to develop novel
strong-coupling theories for strongly correlated Fermi or Bose systems.
\begin{acknowledgments}
This research was supported by the Australian Research Council's (ARC)
Discovery Program, Grants Nos. DP240101590 (H.H.), FT230100229 (J.W.),
and DP240100248 (X.-J.L.).
\end{acknowledgments}

\appendix

\section{The antisymmetrization of $\sum_{j}A_{j}$}

Here, we wish to show that the following function $Z(q_{1},\cdots,q_{n})$
is antisymmetric with respect to the exchange of any two momenta in
the set $\{q_{l}\}$ ($l=1,\cdots,n$):
\begin{equation}
Z\equiv\sum_{i}\left(-1\right)^{n-i}z\left(q_{1},\cdots,q_{i-1},q_{i+1},\cdots,q_{n}\right),\label{eq: Zq1qn}
\end{equation}
where the function $z$ itself is already an antisymmetric function
of its $n-1$ arguments. 

Let us suppose that we exchange the two arguments $q_{\alpha}$ and
$q_{\beta}$ in $Z(q_{1},\cdots,q_{n})$, where $\alpha<\beta$ and
$\alpha,\beta=1,\cdots,n$. If the $i$-th sub-term $(-1)^{n-i}z(q_{1},\cdots,q_{i-1},q_{i+1},\cdots,q_{n})$
involve both $q_{\alpha}$ and $q_{\beta}$, then it is already antisymmetrized.
Thus, we only need to consider two sub-terms, 
\begin{eqnarray}
D_{\alpha} & \equiv & \left(-1\right){}^{n-\alpha}z\left(q_{1},\cdots,q_{\alpha-1},q_{\alpha+1},\cdots,q_{n}\right),\\
D_{\beta} & \equiv & \left(-1\right){}^{n-\beta}z\left(q_{1},\cdots,q_{\beta-1},q_{\beta+1},\cdots,q_{n}\right).
\end{eqnarray}
 Upon the exchange of the two arguments $q_{\alpha}$ and $q_{\beta}$,
in $D_{\alpha}$ the argument $q_{\beta}$ becomes $q_{\alpha}$ and
we need to transport this $q_{\alpha}$ all the way backward to the
position $\alpha$. During this transportation, a sign $(-1)^{\beta-\alpha-1}$
appears due to the antisymmetrization of the function $z$. Therefore,
we obtain $D_{\alpha}\rightarrow D'_{\alpha}=-D_{\beta}$. In $D_{\beta}$,
the argument $q_{\alpha}$ becomes $q_{\beta}$ and similarly we need
to transport this $q_{\beta}$ all the way forward to the position
$\beta$. As a result, $D_{\beta}\rightarrow D'_{\beta}=-D_{\alpha}$.
Therefore, we conclude
\begin{equation}
D'_{\alpha}+D'_{\beta}=-\left(D_{\alpha}+D_{\beta}\right).
\end{equation}
Putting all the sub-terms together, we observe that $Z(q_{1},\cdots,q_{n})$
is indeed an antisymmetric function.

In Eq. (\ref{eq: Zq1qn}), let us take $i=n-j+1$ and $z(q_{1},\cdots,q_{i-1},q_{i+1},\cdots,q_{n})=(-1)^{j-1}A_{n-j+1}$,
we see immediately that $Z=\sum_{j}A_{j}$. Hence, $\sum_{j}A_{j}$
is an antisymmetric function upon the exchange of any two momenta
in $\{q_{l}\}$ ($l=1,\cdots,n$).

\section{Two rules on the Matsubara frequency summation}

In this appendix, we establish the two rules on the Matsubara frequency
summation. Let us first consider the second rule Eq. (\ref{eq: MatFreqSumRule2}),
and apply it to the vertex function $\Gamma_{2}(k;p,q)$ or its cousin
$\gamma_{2}(p,q)$, as an example. 

Roughly speaking, the vertex function $\Gamma_{2}(k;p,q)$ represents
the Green function of a molecule, i.e., a quasi-bound state of two
fermions with unlike spin (i.e., a spin-up fermion and the impurity).
The vertex function does not have singularity at negative energy,
if the molecular state is not the ground state, a situation that we
will focus on. $\gamma_{2}(p,q)$ would have the similar behavior.
Let us write down $\gamma_{2}(p,q)$ in its spectral representation,
for example ($q=\{\mathbf{q},i\omega_{q}\}$),
\begin{equation}
\gamma_{2}\left(p,q\right)=\int_{-\infty}^{+\infty}\frac{d\omega'}{\pi}\left[\frac{-\textrm{Im}\gamma_{2}\left(p,\left\{ \mathbf{q},\omega'\right\} \right)}{i\omega_{q}-\omega'}\right].
\end{equation}
A summation over the Matsubara frequency leads to the density of molecules,
$\int[d\omega'/\pi][-\textrm{Im}\gamma_{2}(p,\{\mathbf{q},\omega'\})]f(\omega')$,
which should be vanishingly small in the thermodynamic limit. Here,
$f(\omega')\equiv1/(e^{\omega'/k_{B}T}+1)$ is the Fermi-Dirac distribution
function. We do not worry about the negative energy, since $\gamma_{2}(p,\{q,\omega'\})$
is analytic on the half-plane $\textrm{Re}\omega'<0$, so its imaginary
part vanishes there. However, on the other half-plane $\textrm{Re}\omega'>0$,
the vanishing density means we should view the Fermi distribution
function $f(\omega')$ as an infinitesimal, i.e., $f(\omega')=1/V\rightarrow0$,
where $V\rightarrow\infty$ is the volume of the whole system. 

Let us now explicitly integrate out the Matsubara frequency in the
expression for the self-energy, Eq. (\ref{eq: SelfEnergy}),
\begin{eqnarray}
\Sigma\left(p\right) & = & \sum_{\mathbf{q}}k_{B}T\sum_{i\omega_{q}}\int\frac{d\omega'}{\pi}\frac{-\textrm{Im}\gamma_{2}\left(p,\left\{ \mathbf{q},\omega'\right\} \right)}{\left(i\omega_{q}-\xi_{\mathbf{q}}\right)\left(i\omega_{q}-\omega'\right)},\\
 & = & \sum_{\mathbf{q}}\int\frac{d\omega'}{\pi}\left[-\textrm{Im}\gamma_{2}\left(p,\left\{ \mathbf{q},\omega'\right\} \right)\right]\frac{f\left(\xi_{\mathbf{q}}\right)}{\xi_{\mathbf{q}}-\omega'},\\
 & = & \sum_{\mathbf{q}}f\left(\xi_{\mathbf{q}}\right)\gamma_{2}\left(p,\left\{ \mathbf{q},\xi_{\mathbf{q}}\right\} \right),
\end{eqnarray}
where in the second line we have taken $f(\omega')=0$ as we emphasized
earlier. This result can be easily recognized from the associated
diagram and can be generalized as the second rule Eq. (\ref{eq: MatFreqSumRule2}).
The integration is for a Fermi loop that winds back a fermion line,
on which the fermionic Matsubara frequency is to be summed. We can
simply replace the backward Green function by a Fermi distribution
function with an on-shell energy $\xi_{\mathbf{q}}=\hbar^{2}\mathbf{q}^{2}/(2m)-\mu$,
where $\mu$ is the chemical potential of the Fermi sea. The energy
$\omega'$ in $\gamma_{2}(p,\{\mathbf{q},\omega'\})$ should also
then be replaced by the on-shell value $\xi_{\mathbf{q}}$.

How about the first rule on the summation over the forward momentum
$k=\{\mathbf{k},i\omega_{k}\rightarrow\omega'+i0^{+}\}$, i.e., $\sum_{k}G_{0\uparrow}(k)G_{0\downarrow}(p+q-k)\Gamma_{3}(k;p,qq')$,
where $q$ and $q'$ may take on-shell values? This summation appears
in the last term of the right-hand-side of Eq (\ref{eq: Gamma2})
and for clarity we have rename $k'$ as $k$. It is easy to see that
the pole of the bare impurity Green function $G_{0\downarrow}(p+q-k)$
occurs at $\omega'=\omega+\xi_{\mathbf{q}}-\varepsilon_{\mathbf{p}+\mathbf{q}-\mathbf{k}}^{I}+\mu_{\downarrow}$,
where $\mu_{\downarrow}\rightarrow-\infty$ is the impurity chemical
potential. Therefore, we have $\textrm{Re}\omega'<0$ and $G_{0\downarrow}(p+q-k)$
is analytic on the half-plane $\textrm{Re}\omega'>0$. We would assume
that for the argument $k$, $T_{3}(k;p,qq')$ is also analytic on
the half-plane $\textrm{Re}\omega'>0$ and may then write $T_{3}(k;p,qq')$
in the spectral representation. By repeating the similar reasons for
$\gamma_{2}(p,q)$, we find that the vanishing density related to
$T_{3}(k;p,qq')$ (i.e., the density of trimer) implies that we should
take the Fermi distribution function $f(-\omega')$ as an infinitesimal
on the half-plane $\textrm{Re}\omega'<0$, when we handle the forward,
particle-like four-momentum $k$.

Now, let us denote collectively $P(k)=G_{0\downarrow}(p+q-k)\Gamma_{3}(k;p,qq')$,
in which the other arguments other than $k$ are made implicit. As
$P(k)$ is analytic on the half-plane $\textrm{Re}\omega'>0$, we
find that $I=\sum_{k}G_{0\uparrow}(k)P(k)$,
\begin{eqnarray}
I & = & \sum_{\mathbf{k}}k_{B}T\sum_{i\omega_{k}}\int\frac{d\omega'}{\pi}\frac{-\textrm{Im}P\left(\left\{ \mathbf{k},\omega'\right\} \right)}{\left(i\omega_{k}-\xi_{\mathbf{k}}\right)\left(i\omega_{k}-\omega'\right)},\\
 & = & \sum_{\mathbf{k}}\int\frac{d\omega'}{\pi}\left[-\textrm{Im}P\left(\left\{ \mathbf{k},\omega'\right\} \right)\right]\frac{-f\left(-\xi_{\mathbf{k}}\right)}{\xi_{\mathbf{k}}-\omega'},\\
 & = & -\sum_{\mathbf{k}}f\left(-\xi_{\mathbf{k}}\right)P\left(\left\{ \mathbf{k},\xi_{\mathbf{k}}\right\} \right),
\end{eqnarray}
where in the second line we have taken $f\left(-\omega'\right)=0$
on the half-plane $\textrm{Re}\omega'<0$, on which $-\textrm{Im}P\left(\left\{ \mathbf{k},\omega'\right\} \right)$
may develop poles. Thus, for the summation over the fermionic Matsubara
frequency for the particle-like momentum $k$, we can simply replace
the forward Green function by the Fermi distribution function $-f(-\xi_{\mathbf{k}})$
with the on-shell energy $\xi_{\mathbf{k}}=\hbar^{2}\mathbf{k}^{2}/(2m)-\mu$,
and then take $i\omega_{k}\rightarrow\xi_{\mathbf{k}}$ in $P(k,\cdots)$.
This is exactly the first rule, Eq. (\ref{eq: MatFreqSumRule1}).

\section{Chevy ansatz for molecules}

For very strong attraction, Fermi polarons may become unstable and
turn into tightly bound molecules that are dressed by particle-hole
excitations of the Fermi sea \citep{Punk2009,Mora2009,Combescot2009}.
Here, we would like to list the set of the equations for molecules,
which can be easily derived following Sec. IV. The Chevy ansatz $\left|\Phi\right\rangle $
for molecules with momentum $\mathbf{p}$ takes the form,\begin{widetext}
\begin{eqnarray}
\left|\Phi\right\rangle  & = & \left[\sum_{\mathbf{k_{0}}}\varphi^{\mathbf{k}_{0}}d_{\mathbf{p}-\mathbf{\mathbf{k}_{0}}}^{\dagger}c_{\mathbf{\mathbf{k}_{0}}}^{\dagger}+\frac{1}{2!}\sum_{\mathbf{k}_{0}\mathbf{kq}}\varphi_{\mathbf{q}}^{\mathbf{\mathbf{k}_{0}}\mathbf{k}}d_{\mathbf{p}-\mathbf{\mathbf{k}_{0}}+\mathbf{q}-\mathbf{k}}^{\dagger}c_{\mathbf{\mathbf{k}_{0}}}^{\dagger}c_{\mathbf{k}}^{\dagger}c_{\mathbf{q}}+\cdots\right]\left|\textrm{FS}\right\rangle _{N-1},\\
 & = & \sum_{n=0}^{\infty}\frac{1}{n!\left(n+1\right)!}\sum_{\mathbf{k}_{0}\mathbf{k}_{1}\cdots\mathbf{k}_{n}\mathbf{q}_{1}\cdots\mathbf{q}_{n}}\varphi_{\mathbf{q}_{1}\cdots\mathbf{q}_{n}}^{\mathbf{k}_{0}\mathbf{k}_{1}\cdots\mathbf{k}_{n}}d_{\mathbf{p}-\mathbf{k}_{0}+(\mathbf{q}_{1}+\cdots+\mathbf{q}_{n})-(\mathbf{k}_{1}+\cdots+\mathbf{k}_{n})}^{\dagger}c_{\mathbf{k}_{0}}^{\dagger}c_{\mathbf{k}_{1}}^{\dagger}\cdots c_{\mathbf{k}_{n}}^{\dagger}c_{\mathbf{q}_{n}}\cdots c_{\mathbf{q}_{1}}\left|\textrm{FS}\right\rangle _{N-1},\label{eq: ChevyAnsatzMolecules}
\end{eqnarray}
where $\left|\textrm{FS}\right\rangle _{N-1}$ describes a thermal
Fermi sea at finite temperature with $N-1$ fermions. We find that
the coefficients $\varphi_{\mathbf{q}_{1}\cdots\mathbf{q}_{n}}^{\mathbf{k}_{0}\mathbf{k}_{1}\cdots\mathbf{k}_{n}}$
satisfy the following coupled equations, 

\begin{eqnarray}
-\mathcal{E}_{\mathbf{p};\mathbf{k}_{0}\{\mathbf{k}\};\{\mathbf{q}\}}^{(n)}\varphi_{\mathbf{q}_{1}\cdots\mathbf{q}_{n}}^{\mathbf{k}_{0}\mathbf{k}_{1}\cdots\mathbf{k}_{n}} & = & U\sum_{j=1,\cdots,n}^{i=0,\cdots,n}\left(-1\right)^{i+j-1}\varphi_{\mathbf{q}_{1}\cdots\mathbf{q}_{n-j}\mathbf{q}_{n-j+2}\cdots\mathbf{q}_{n}}^{\mathbf{k}_{0}\mathbf{k}_{1}\cdots\mathbf{k}_{n-i-1}\mathbf{k}_{n-i+1}\cdots\mathbf{k}_{n}}+U\sum_{\mathbf{K}}\left(\varphi_{\mathbf{q}_{1}\mathbf{q}_{2}\cdots\mathbf{q}_{n}}^{\mathbf{K}\mathbf{k}_{1}\cdots\mathbf{k}_{n}}+\cdots+\varphi_{\mathbf{q}_{1}\mathbf{q}_{2}\cdots\mathbf{q}_{n}}^{\mathbf{k}_{0}\mathbf{k}_{1}\cdots\mathbf{k}_{n-1}\mathbf{K}}\right)f\left(-\xi_{\mathbf{K}}\right)\nonumber \\
 &  & -U\sum_{\mathbf{Q}}\left(\varphi_{\mathbf{Q}\mathbf{q}_{2}\cdots\mathbf{q}_{n}}^{\mathbf{k}_{0}\mathbf{k}_{1}\cdots\mathbf{k}_{n}}+\cdots+\varphi_{\mathbf{q}_{1}\cdots\mathbf{q}_{n-1}\mathbf{Q}}^{\mathbf{k}_{0}\mathbf{k}_{1}\cdots\mathbf{k}_{n}}\right)f\left(\xi_{\mathbf{Q}}\right)+U\sum_{\mathbf{KQ}}\varphi_{\mathbf{q}_{1}\mathbf{q}_{2}\cdots\mathbf{q}_{n}\mathbf{Q}}^{\mathbf{k}_{0}\mathbf{k}_{1}\cdots\mathbf{k}_{n}\mathbf{K}}f\left(-\xi_{\mathbf{K}}\right)f\left(\xi_{\mathbf{Q}}\right),\label{eq:ChevyAnsatzMoleculesSolution}
\end{eqnarray}
where $\mathcal{E}_{\mathbf{p};\mathbf{k}_{0}\{\mathbf{k}\};\{\mathbf{q}\}}^{(n)}\equiv\left(-E+E_{\textrm{FS},N-1}+\nu U\right)+\varepsilon_{\mathbf{p}-\mathbf{k}_{0}-\mathbf{P}_{\vec{\kappa}_{n}}}^{I}+E_{\vec{\kappa}_{n}}+\varepsilon_{\mathbf{k}_{0}}$
and $E_{\textrm{FS},N-1}$ is the energy of the Fermi sea with $N-1$
fermions.\end{widetext}

\end{document}